\documentclass[aps,twocolumn,showkeys]{revtex4-2}
\usepackage[latin1,utf8]{inputenc} %paquete para las tildes
\usepackage{amssymb}
\usepackage{amsfonts} %paquete para escribir letras como el cuerpo de los reales
\usepackage{graphicx} %paquete grafico
\usepackage{subcaption}
\usepackage{svg}
\usepackage{float}% floating figures [H]
\usepackage{amsthm} %paquete matematico
\usepackage{amsmath} %paquete matematico
\usepackage{bm}%bold lowercase math
\usepackage{epsfig}
\usepackage{hyperref}
\usepackage{color}
\usepackage{bbold}
 \usepackage{bbm}
\usepackage[T1]{fontenc}
\usepackage{young}
\usepackage{youngtab}
\usepackage{xcolor}
\usepackage{braket}
\usepackage{lipsum}
\usepackage{comment}
\def\ic{\mathrm{i}}

\def \bc {\begin{center}}
\def \ec {\end{center}}
\def \bi {\begin{itemize}}
\def \ei {\end{itemize}}

\def \ba {\begin{array}}
\def \ea {\end{array}}

\def \bea {\begin{eqnarray}}
\def \eea {\end{eqnarray}}

\def \be {\begin{equation}}
\def \ee {\end{equation}}

\newcommand{\la}{\langle}
\newcommand{\ra}{\rangle}

\def\bc {\bar{\beta}}

\def\nb{{\vec{n}}}
\def\zb{\bm{z}}

\def\rmu{\mathrm{U}}

\def\2cat{{\scriptstyle\mathrm{2CAT}}}
\def\3cat{{\scriptstyle\mathrm{3CAT}}}

\definecolor{orange2}{RGB}{214,186,61}
\definecolor{darkbrown}{RGB}{105, 54, 0}

\begin{document}

% \title{Phase space analysis of Lipkin-Meshkov-Glick D-level atom critical models: \\ Husimi function, Inverse Participation Ratio and Wehrl Entropy \\
% Phase space analysis of the Lipkin-Meshkov-Glick 3-level model using localization measures of parity adapted U($D$)-spin coherent states\\
%\tcb{Localization measures of parity adapted U($D$)-spin coherent states applied to the phase space analysis of the Lipkin-Meshkov-Glick 3-level model}
%\title{Excited-state quantum phase transitions in 3-level atom models}
% \title{Excited-state quantum phase transitions in a 3-level Lipkin model}
\title{Excited-state quantum phase transitions and chaos in a three-level Lipkin model}

\author{Alberto Mayorgas}
\email{alberto.mayorgasreyes@ceu.es}
\affiliation{Escuela Polit\'ecnica Superior, Universidad CEU Fernando III, CEU Universities,
	Glorieta Cardenal Herrera Oria, 41930 Sevilla, Spain}
\author{Pedro P\'erez-Fern\'andez}
\email{pedropf@us.es}
\affiliation{Departamento de F\'isica Aplicada III, Escuela T\'ecnica Superior de Ingenier\'ia, Universidad de Sevilla, 41092 Sevilla, Spain\\
 Multidisciplinary Unit for Energy Science, Universidad de Sevilla, 41080 Sevilla, Spain}
\author{\'Alvaro S\'aiz}
\email{asaiz@us.es}
\affiliation{Departamento de F\'isica Aplicada III, Escuela T\'ecnica Superior de Ingenier\'ia, Universidad de Sevilla, 41092 Sevilla, Spain}
\affiliation{Departamento de F\'isica At\'omica, Molecular y Nuclear, Facultad de F\'isica, Universidad de Sevilla, Apartado 1065, 41080 Sevilla, Spain}
\author{Jos\'e Miguel Arias}
\email{ariasc@us.es}
\affiliation{Departamento de F\'isica At\'omica, Molecular y Nuclear, Facultad de F\'isica, Universidad de Sevilla, Apartado 1065, 41080 Sevilla, Spain\\
Multidisciplinary Unit for Energy Science, Universidad de Sevilla, 41080 Sevilla, Spain}
%\author{Julio Guerrero}
%\email{jguerrer@ujaen.es}
%\affiliation{Department of Mathematics, University of Jaen, Campus Las Lagunillas s/n, 23071 Jaen, Spain}
%\affiliation{Institute Carlos I of Theoretical and Computational Physics (iC1), University of  Granada,
%	Fuentenueva s/n, 18071 Granada, Spain}
%\author{Manuel Calixto}
%\email{calixto@ugr.es}
%\affiliation{Department of Applied Mathematics, University of  Granada,
%	Fuentenueva s/n, 18071 Granada, Spain}
%\affiliation{Institute Carlos I of Theoretical and Computational Physics (iC1), University of  Granada,
%	Fuentenueva s/n, 18071 Granada, Spain}

\date{\today}

\begin{abstract}
	\vspace{1cm}
	\section*{Abstract}

Excited-state quantum phase transitions (ESQPTs) have been extensively studied in two-level models, but their characterization remains challenging in systems displaying mixed regular and chaotic dynamics. In this work, we investigate ESQPTs within the three-level Lipkin–Meshkov–Glick model, where an enlarged Hilbert space and multiple separatrices give rise to rich spectral structures strongly influenced by chaos. To investigate the different dynamical regions, we have calculated Poincaré sections and Peres lattices. In addition, by combining chaos-sensitive measures with standard ESQPT diagnostics, we provide a static analysis of ESQPT signatures in this model and establish a robust framework for future studies of its dynamical behavior. The degree of chaos and the Kullback-Leibler divergence are found to be very effective chaos-sensitive measures, which are complementary to ESQPT diagnostics such as the mean field limit and the participation ratio. Hence we provide a standard framework to work with ESQPTs in chaotic three-level systems. 
	%\textcolor{gray}{\lipsum[1]}
\end{abstract}

% \noindent \textbf{PACS:}
% %75.10.Jm % Heisenberg model, Quantized spin models
% 73.43.-f,   %Fractional quantum Hall effect 
% 03.65.Fd, %Algebraic methods
% 03.65.Ge,   %Harmonic oscillators
%  
%PACS. 03.65.Ud Entanglement and quantum nonlocality (e.g. EPR paradox, Bell's inequalities, GHZ states, etc.) - 03.67.Lx Quantum computation - 75.10.Jm Quantized spin models
% 
% 
% \noindent \textbf{MSC:}
% 81V70, %Many-body theory; quantum Hall effect
% % 81R30, %Coherent states (See also 22E45); squeezed states (See also 81V80)
% 81Rxx, %Groups and algebras in quantum theory
% %14M15 %Grassmannians, Schubert varieties, flag manifolds (See also 32M10, 51M35)
% 
% 

%\keywords{Quantum phase transitions, many-body systems, symmetric quDits, coherent states, entanglement entropies.  }

\maketitle

%\begin{figure}[h]
%	\centering
%	\includegraphics[width=0.5\textwidth]{Figures/EnergyLevels_Num_LMGU3.pdf}
%	\caption{}
%	\label{}
%\end{figure}

\section{Introduction}
Matter can exist in distinct structural states, or phases, each characterized by specific physical properties. Transitions between phases are accompanied by abrupt changes in these properties. The study of phase transitions and critical phenomena is of fundamental importance because of the emergence of universal behavior largely independent of microscopic details  \cite{PT_textbook,PT_textbook2}.

A phase transition is typically driven by a control parameter—such as temperature or another relevant variable—and characterized by an order parameter that vanishes in one phase and becomes finite in another. The evolution of this order parameter near the transition determines the classification of the phase transition  \cite{PT_textbook,PT_textbook2}.

The concept of phase transitions, originally developed for macroscopic systems, extends to the quantum domain, giving rise to quantum phase transitions (QPTs) \cite{Sachdev2011, Vojta_2003, Carr2010}. These transitions occur when a control parameter reaches a critical value and have been characterized through various static and dynamic observables. Abrupt variations in ground-state properties define ground-state quantum phase transitions (GSQPTs), whereas singularities in the level density at specific excitation energies mark excited-state quantum phase transitions (ESQPTs) \cite{CejnarReview}. Such singularities emerge in the thermodynamic limit, where the number of particles tends to infinity, although finite systems display precursor signatures of this critical behavior. Both GSQPTs and ESQPTs have been investigated in numerous two-level models (e.g., Refs. \cite{CejnarReview,STRANSKY201473,PhysRevC.99.064323,PhysRevE.107.064134,PhysRevE.106.044125,PhysRevA.83.033802,PhysRevE.104.064116,PhysRevE.83.046208,PhysRevC.93.044302}).

The objective of this work is to analyze GSQPTs and ESQPTs in a more complex system that exhibits regular, chaotic, and mixed dynamical regimes across different excitation energies. To this end, we employ the three-level Lipkin–Meshkov–Glick (LMG) model as a paradigmatic example  \cite{Meredith,KusLipkin}, which is the extension of the two-level LMG \cite{lipkin1,lipkin2}. Comparing for a fixed number of particles $N$, the inclusion of a third level in the LMG model increases the Hilbert-space dimensionality and the number of level crossings, producing more stationary points of the energy surface and more candidates to be separatrices.
%multiple separatrices that partition the spectrum into regions with varying degrees of chaos. 
Characterizing ESQPT under such conditions is challenging \cite{PhysRevE.83.046208},  as the discontinuities of the first derivative of the level density, characterizing ESQPT in models with two degrees of freedom \cite{CejnarReview}, 
interfere with the avoided crossings typical of chaotic dynamics \cite{Stockmann1999}.

 Despite a general connection between ESQPT and onset of chaos does not exist \cite{STRANSKY201557,Cejnar_2015},
we establish a particular link between both concepts in the 3-level LMG in the strong interaction limit. For this purpose, we analyze three different dynamical regimes (quasi-integrable, chaotic, and quasi-chaotic) by computing Poincaré sections \cite{PoincareSect} in phase space and Peres lattices \cite{Peres1984} for selected observables. In addition, to characterize the different regions statically we employ measures of chaos and fluctuation—including the nearest-neighbor spacing distribution (NNSD), the degree of chaos $\eta$, the Kullback-Leibler divergence, and the overlap distribution \cite{Meredith,PhysRevE.83.046208,PhysRevA.101.053604,Kullback-Liebler}—together with conventional ESQPT indicators such as the adjacent-level energy gap and the participation ratio. Here, we focus on the static characterization of ESQPTs in the three-level LMG model, leaving the analysis of dynamical aspects for future work.

The paper is organized as follows. Section \ref{sec2} introduces the three-level LMG model, the associated parity operators, and the Hilbert space structure. Section \ref{sec3} presents the classical limit obtained through an analysis of the energy surface and phase diagram of the system. Section \ref{sec4} discusses the effects of ESQPTs on the regular-to-chaotic transition, including comparisons between Peres lattices and Poincaré sections, NNSD analysis, and the study of the Kullback-Leibler divergence, participation ratio, and overlap distribution. Finally, our conclusions are summarized in Sec. \ref{conclusion}.

\section{Model}\label{sec2}

%\subsection{3-level Lipkin-Meshkov-Glick Hamiltonian}

Many-body Hamiltonians of $N$ identical particles in a $D$-level system and with infinite-range interactions are usually defined in terms of the U($D$) Lie algebra and its associated collective generators $S_{ij}$, with $i,j=0,1,\ldots,D-1$, 
that fulfill the U($D$) commutation relations 
\be
\left[S_{{ij}},S_{{kl}}\right]=\delta_{{jk}} S_{{il}} -\delta_{{il}} S_{{kj}}.\label{commurel}
\ee

It is well known that any arbitrary Lie algebra (as U($D$)) can be mapped to an oscillator (bosonic) representation using the Jordan-Schwinger transformation. Usually, a realization in terms of bilinear products of boson creation $b^\dag_i$ and annihilation $b_j$ operators is employed,
\be
S_{ij}=b^\dag_i b_j~ ;~ i,j=0,1,2,\dots,D-1 . \label{UDgen}
\ee
For example, in the case of two-level systems $D=2$ \cite{PhysRevB.74.104118,Romera2014,PhysRevC.79.014301,DUZZIONI2004322,PhysRevC.110.044318,PhysRevA.83.062327,10.1063/1.528042}, the SU(2) angular momentum collective operators are $(J_x,J_y,J_z)=\sum_{\mu=1}^{N}(\sigma_x^{\mu},\sigma_y^{\mu},\sigma_z^{\mu})$, with $\mu=1,\ldots,N$ identical particles, and the following identifications are made $J_+=S_{10}, J_-=S_{01}$, $J_z=\frac{1}{2}(S_{11}-S_{00})$. In addition, if the total number of particles is conserved, there is an extra conserved quantity: $C_1=S_{00}+S_{11}$ (the linear Casimir operator of $\rmu(2)$).

\begin{figure}[h]
	\centering
	\includegraphics[width=0.35\textwidth]{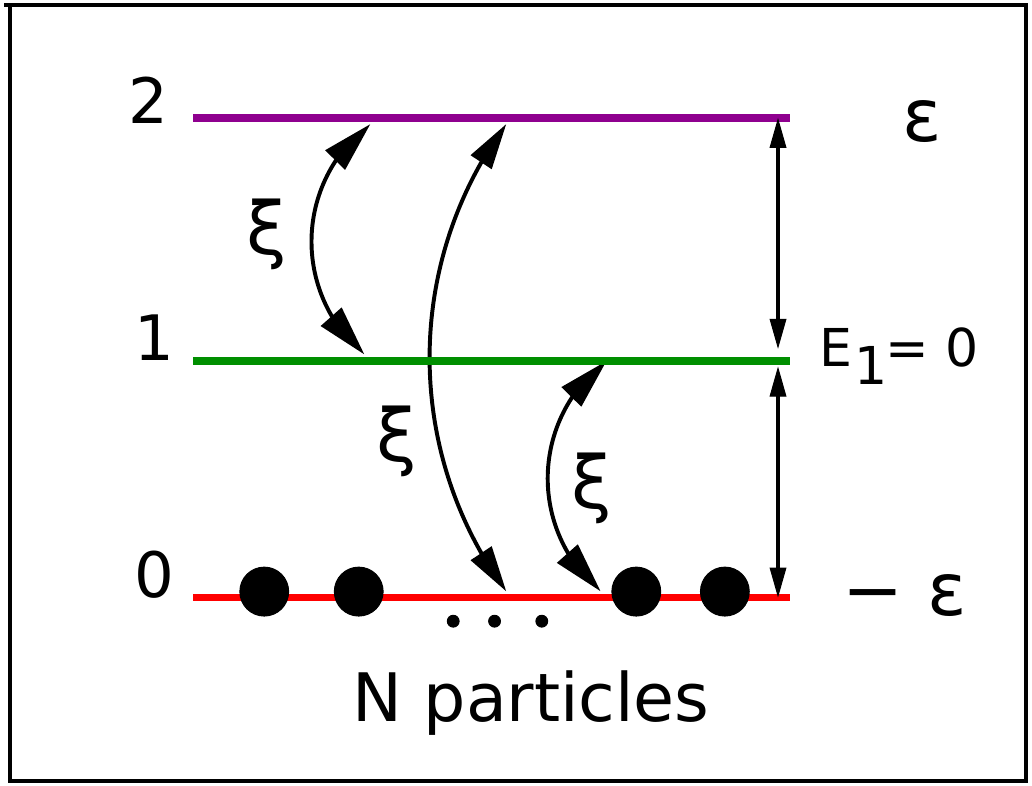}
	\caption{Schematic image of the 3-level Lipkin model used in this work, described by Hamiltonian (\ref{hamU3old}).}
	\label{Fig:Schem3levelLM}
\end{figure}

In this work, the 3-level LMG Hamiltonian with a fixed number of particles is considered
\begin{equation}
    H = \sum_{i=0}^{2} E_i ~ S_{ii}  -    % \frac{1}{2} ~  
    \sum_{i\not=j=0}^{2} \xi_{ij} ~ S_{ij}^2 ~,
\end{equation}
where $E_i$ is the energy of the $i-$level and $\xi_{ij}$ are the strengths of the two-body interaction between levels $i$ and $j$. In this work, we have considered the same strength $\xi=\xi_{ij} ~; ~ \forall i,j$ (control parameter) in all levels for the two-body interaction scattering pairs of particles and equal energy spacing, $\varepsilon$,  between levels, as in previous studies \cite{Meredith,Lopez-Arias1989,Kus,KusLipkin,Casati,Saraceno,nuestroPRE}. We have taken $E_1=0$ as represented in Fig. \ref{Fig:Schem3levelLM}.
%s considered,
%\begin{equation}
%    H = E_0 \left( S_{00} + \varepsilon S_{11} +2 \varepsilon S_{22} \right)  -    % \frac{1}{2} ~  
%    \sum_{i\not=j=0}^{2} \eta_{ij} ~ S_{ij}^2 ~,
%\end{equation}
%where $E_0$ is the $i=0$ energy and $\eta_{ij}$ is the strength of the two-body interaction between levels $i$ and $j$. 
  In addition, the total number of particles $N=S_{00}+S_{11}+S_{22}$ (the linear Casimir operator of $\rmu(3)$) is conserved. %A schematic representation of the model is given in Fig. \ref{Fig:Schem3levelLM}. 
In this situation, and using the U(3) collective operators, the Hamiltonian %(apart from an irrelevant constant factor) 
takes the form,
\begin{equation}
	H=\varepsilon(S_{22}-S_{00}) - \xi ~%\frac{1}{2}
    \sum_{i\not=j=0}^{2} S_{ij}^2.\label{hamU3old}
\end{equation}

%%%%%%%%%%%%%%%%%%%%%%%%%%%%%%%
We can conveniently renormalize one-body interactions $\varepsilon\to \varepsilon/N$ by the total number $N$ of particles, and two-body interactions $\xi \to \xi/[N(N-1)]$ by a factor related to the total number $N(N-1)/2$ of pairs. Furthermore, we reduce the total number of parameters by expressing the Hamiltonian dimensionless, that is dividing both terms in (\ref{hamU3old}) by $\varepsilon$ and setting $\xi/ \varepsilon \rightarrow \lambda$. In addition to all these simplifications, it is convenient to multiply the one-body interactions by $(1-\lambda)$ in order to achieve a compact form of the Hamiltonian $H=(1-\lambda)H_0-\lambda H_1$, with a bounded control parameter $\lambda\in[0,1]$. This reparameterization allows us to characterize chaotic regions of the eigenspectrum and ESQPT for finite values of $\lambda\simeq1$ (see Sec.\ref{Sec:NNSD}), preventing the selection of high values of $\lambda$ as in \cite{Meredith,nuestroPRE}. In summary, the Hamiltonian per particle used in this work is written as
\begin{equation}
	H=\frac{(1-\lambda)}{N}(S_{22}-S_{00})-\frac{\lambda}{N(N-1)}\sum_{i\not=j=0}^{2} S_{ij}^2.\label{hamU3}
\end{equation}
As in the two-level case, the model exhibits parity symmetry. That is, the interaction only scatters pairs of particles, so that the population $S_{kk}$ in each level $k=0,1,2$ conserves the even (+) or odd (-) parity. The operators related to this conservation are $\Pi_k = \exp(\ic \pi S_{kk})$. Hence, four different Hamiltonian independent sectors are identified by $(\Pi_0,\Pi_1,\Pi_2)$,
\begin{align}
	(+,+,+)&, (-,-,+), (-,+,-), (+,-,-)\quad N\text{ even}\label{parityEven}\\
    (-,-,-)&, (+,+,-), (+,-,+), (-,+,+)\quad N\text{ odd}\nonumber
    \end{align}
%\end{equation}
This discrete symmetry corresponds to the finite group $\mathbb{Z}_2\times\mathbb{Z}_2\times\mathbb{Z}_2$, with the constraint 
$\Pi_0\Pi_1\Pi_2=e^{i\pi N}$. The parity symmetry is broken in the thermodynamic limit, so the ground state becomes highly degenerated for the case $D=3$ compared to $D=2$.

\begin{figure}[h]
	\centering

\begin{subfigure}[b]{0.48\textwidth}
	\includegraphics[width=1\textwidth]{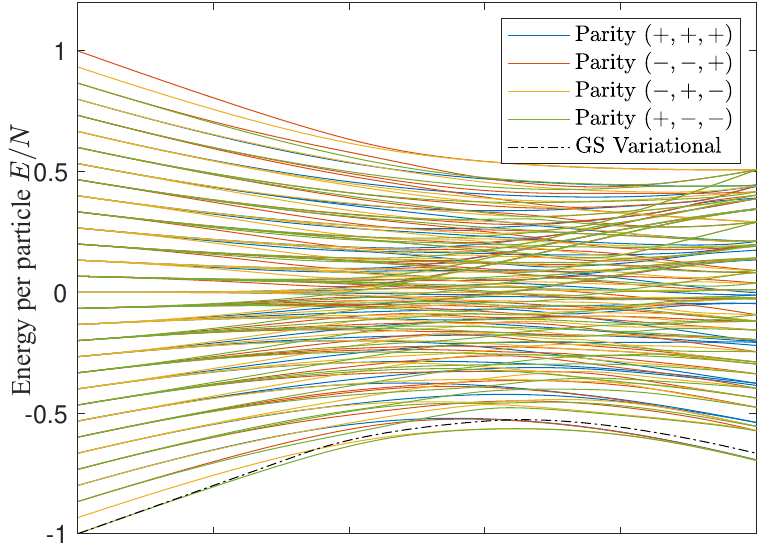}
\end{subfigure}

\medskip % insert a bit of vertical whitespace
\begin{subfigure}[b]{0.48\textwidth}
    \includegraphics[width=1\textwidth]{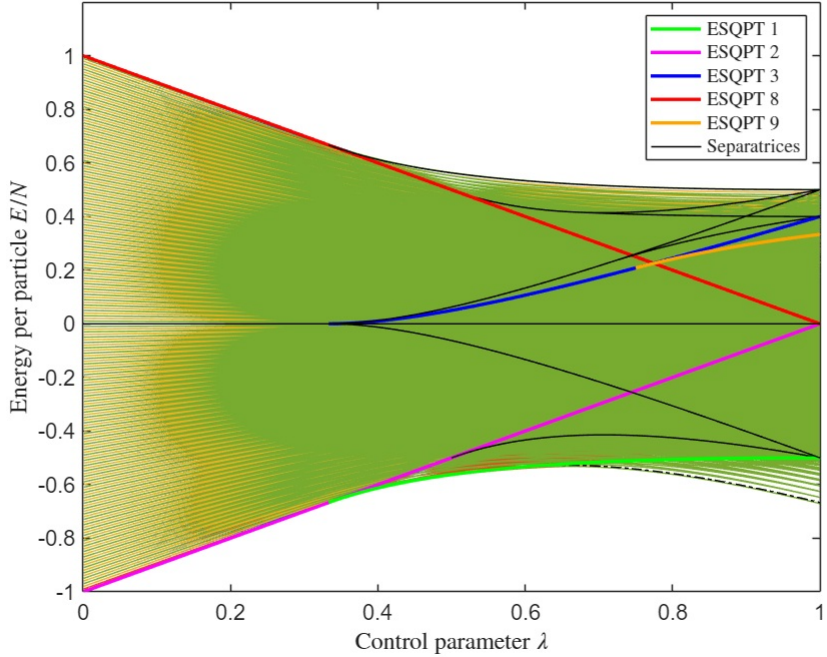}
\end{subfigure}

\caption{Energy per particle $E/N$ of the symmetric representation of the 3-level LMG model as a function of the control parameter $\lambda$, for $N=15$ (top) and $N=100$ (bottom) particles. In both pictures, the thin lines represent the numerical energies, whose colors describe the parity sector in which they have been calculated. The dashed-dotted black line is the variational ground state \eqref{energysym}, whose convergence to the numerical GS increases with $N$. The thick colored lines in the bottom spectrum represent the most relevant ESQPT separatrices  \eqref{Separatrix1}, \eqref{Separatrix2}, while the rest are given in thin black lines \eqref{Separatrices}. See Appendix \ref{App:C} for a discussion about the $\lambda$-domain of the separatrices.}
\label{Fig:Energy100}
\end{figure}

For the sake of simplicity, we consider indistinguishable particles and reduce the Hilbert space dimension from $3^N$ to $d=\tbinom{N+2}{2}=\tfrac{(N+1)(N+2)}{2}$, corresponding to the fully symmetric irreducible representation of U(3) \cite{Barut}. As a consequence, computational complexity is reduced for large $N$. The rest of the mixed permutation symmetry sectors have already been studied in \cite{nuestroPRE}. In the fully symmetric case, the Hilbert space is spanned by the Bose-Einstein-Fock basis  states
\be
|\vec{n}\ra=|n_0,n_1,n_2\ra=
\frac{(b_0^\dag)^{n_0}(b_1^\dag)^{n_1}(b_2^\dag)^{n_2}}{\sqrt{n_0!n_1!n_2!}}|\vec{0}\ra, \label{symmetricbasis}
\ee
where $|\vec{0}\ra$ is the Fock vacuum, and $n_i$ is the occupancy number of level $i$ (the eigenvalue of $S_{ii}$). The total number of particles is restricted by the expression  $n_0+n_1+n_2=N$. 
% In the low-interaction regime $\lambda\ll 1$, the ground state of \eqref{hamU3} is a Bose-Einstein condensate $\tfrac{1}{\sqrt{N!}}(b_0^\dag)^{N}|\vec{0}\ra$ of $N$ particles in the $i=0$ level, which we shall take as a reference level from now on. 
A general symmetric $N$-particle state $\psi$ in the Fock basis is written as $|\psi\ra=\sum\,c_{\vec{n}}|\vec{n}\ra$
where the sum is restricted to $n_0+n_1+n_2=N$. 
Hence the matrix elements of the collective $\rmu(3)$-spin operators \eqref{UDgen} in the Fock basis are
\begin{align}
	&\la\vec{m}|S_{ii}|\vec{n}\ra=n_i ~ \delta_{\vec{m},\vec{n}}\,,\label{Sijmatrix}\\ 
	&\la\vec{m}|S_{ij}|\vec{n}\ra=\sqrt{(n_i+1)~n_j} ~\delta_{m_i,n_{i}+1}\delta_{m_j,n_{j}-1}\prod_{k\not=i\not=j}\delta_{m_k,n_k}\,.\nonumber
\end{align}
With these matrix elements, the construction of the Hamiltonian per particle (\ref{hamU3}) matrix and its diagonalization are easy and provide the complete energy spectrum for a given number of particles, $N$. In Figure \ref{Fig:Energy100}, the exact numerical energy densities (energy per particle $E/N$) results are plotted against the control parameter, $\lambda$, for $N=15$ particles, in the top panel, and for $N=100$ particles, in the bottom panel. In the diagonalization, the four previously mentioned parity sectors are separated. In this figure, regions around the center of the plot present energy clusters that create an increase of the energy level density typical of an ESQPT. However, the complexity of the spectrum level crossings makes the classification and analysis of these regions difficult. A hint to make this analysis comes from the mean field study valid in the thermodynamic limit, $N \to \infty$. All the diagonalization data is available at the public repository \cite{zenodo_diagonalization-LMGU3N100}.

%%%%%%%%%%%%%%%%%%%%%%%%%%%%%%%%%%%%%%%%%%%%%%%%%%%%%%%%%%%%%%%%%%%%%%%%%%%%%%%%%%%%

\section{Mean field: energy surface and phase diagram}\label{sec3}

Coherent states are used to reproduce the ground-state energy of Hamiltonian models in the limit $N\to\infty$, called the semiclassical or the thermodynamic limit. Therefore, they are often called semiclassical or variational states. $\rmu(3)$-spin coherent states (3SCSs for brevity) are defined as
\begin{equation}
	|\zb\ra=\frac{1}{\sqrt{N!}}\left(
	\frac{b_0^\dag+z_1b_1^\dag+z_2b_2^\dag}{\sqrt{1+|z_1|^2+|z_{2}|^2}}\right)^{N}|\vec{0}\ra,\label{cohD}
\end{equation}
where $z_1,z_2\in\mathbb{C}$ are variational parameters, and the coherent state is associated with $N$ particles in total. The Appendix \ref{statesymmatsec} is devoted to studying 3SCSs in detail.

Using the 3SCS expectation value of the Hamiltonian per particle \eqref{hamU3}, and taking the thermodynamic limit, we define the energy surface as $E(\zb,\lambda)=\lim_{N\to\infty}\langle \zb|H|\zb\rangle$. With the linear $\la \zb|S_{ij}|\zb\ra$ and quadratic $\la \zb|S_{ij}S_{kl}|\zb\ra$ matrix elements given in \eqref{CSEV}, the explicit form for the energy surface is,
\begin{align}
    \label{enersym}
E(\zb,\lambda)= &\,(1-\lambda)\frac{ |z_2|^2-1}{
	1+|z_1|^2+|z_2|^2}\nonumber \\ 
&-\lambda\frac{ z_1 ^2 \left(\bar{z}_2^2+1\right)+z_2^2+\mathrm{c.c.}
}{\left(1+|z_1|^2+|z_2|^2\right)^2}.
\end{align}
The parity symmetry $z_1\to-z_1$, $z_2\to-z_2$ is inherited from the Hamiltonian \eqref{hamU3}.

The ground-state energy corresponds to the variational minimum $E_{0}(\lambda)=\mathrm{min}_{z_1,z_2\in \mathbb{C}}E(z_1,z_2,\lambda)$, which is attained at the critical (real) coherent state parameters
% Stationary_points_LMG_U3.nb
\begin{eqnarray}
z_{1\pm}^{(0)}(\lambda)&=&\pm\left\{\begin{array}{lll}
	0, && 0\leq \lambda \leq \frac{1 }{3}, \\
	\sqrt{\frac{3\lambda- 1 }{ \lambda +1 }}, && \frac{1 }{3}\leq \lambda \leq \frac{3  }{5}, \\
	\sqrt{\frac{2\lambda }{-\lambda +3  }}, && \lambda \geq \frac{3  }{5},
\end{array}\right.\nonumber\\
z_{2\pm}^{(0)}(\lambda)&=&\pm\left\{\begin{array}{lll}
	0, & & 0\leq \lambda \leq  \frac{3  }{5}, \\
	\sqrt{\frac{5 \lambda -3 }{-\lambda +3  }}, & & \lambda \geq \frac{3  }{5}. \end{array}\right. \label{critalphabeta0}
\end{eqnarray}
% \textcolor{blue}{In the 1st phase, $z_{10},z_{20}=0$, so it only exists the symmetric 3CAT$_{(0,0)}$. In the 2nd phase $z_{20}=0$, so there are 2 non null cats 3CAT$_{(0,0)}$, 3CAT$_{(1,0)}$. The second cat is candidate to be 1st excited state,\textcolor{red}{ future work}. In the 3rd phase all the 4 cats are present. }
Inserting \eqref{critalphabeta0} into  \eqref{enersym} gives the ground-state energy per particle in the thermodynamic limit 
\be
E_{0}(\lambda)=\left\{\begin{array}{lllr}
	-1+\lambda,  && 0\leq \lambda \leq \frac{1 }{3}, & \mathrm{(I)}\\
	-\frac{(1+ \lambda )^2}{8 \lambda }, && \frac{1 }{3}\leq \lambda \leq \frac{3 }{5}, &  \mathrm{(II)} \\
	1-\frac{1}{2\lambda}-\frac{7\lambda}{6}, & &\lambda \geq \frac{3  }{5}. &  \mathrm{(III)}\end{array}\right.\label{energysym}
\ee

This energy is plotted in Figure \ref{Fig:Energy100} with a dashed-dotted line under the name GS-variational. As we already anticipated, the ground state is degenerated, since there are four different 3SCSs $|z_{1\pm}^{(0)},z_{2\pm}^{(0)}\ra$ with the same energy \eqref{energysym} in the thermodynamic limit $N\to\infty$. For real systems, there are finite-size effects that slightly modify the ground state, but in Figure \ref{Fig:Energy100} is observed that the exact numerical calculation follows closely the mean field result. Naturally, the larger $N$ is, the better the agreement (see the bottom panel). According to the mean field calculation for the ground-state energy, there are three different phases (I, II and III) separated by two second-order QPTs (according to Ehrenfest's classification) occurring at critical points  $\lambda^{(0)}_{\mathrm{I}\leftrightarrow\mathrm{II}}=1/3$ and $\lambda^{(0)}_{\mathrm{II}\leftrightarrow\mathrm{III}}=3/5$, respectively, at which the second derivative of $E_0(\lambda)$ is discontinuous. 

In the numerical calculation, the complete energy spectrum is obtained, Figure \ref{Fig:Energy100}, and the mean field calculation might help to understand the different structures observed in the excitation spectrum. These are related to the stationary points that appear in the energy surface formula \eqref{enersym}. These are usually called ESQPT separatrices and can be computed as the solutions of the equation $\nabla_{\zb} E(\zb,\lambda)=\mathbf{0}$ \cite{Stransky16, CejnarReview}. Inserting the stationary points into the energy surface formula \eqref{enersym}, we obtain 14 separatrices that are given explicitly in Appendix \ref{App:B} and are plotted in Figure \ref{Fig:Energy100} (bottom panel) as black lines. Some of them are especially important for studying ESQPTs. The first of them is defined by the extension of the GS state energies \eqref{energysym} of phase II and I to larger values of the control parameter, that is,
%Following the notation of
%https://doi.org/10.1103/PhysRevE.106.044125
% Computed in  Stationary_points_LMG_U3.nb
\begin{align}
	\begin{array}{llllr}
    f_1(\lambda)=-\frac{(1+ \lambda )^2}{8 \lambda }, && \forall \lambda \geq \frac{1 }{3}, &&  \mathrm{(ESQPT 1, green)}\\
		f_2(\lambda)=-1+\lambda.  && \forall \lambda \geq 0, && \mathrm{(ESQPT 2, magenta)}\\
	\end{array}\label{Separatrix1}
\end{align}
These are graphically presented by the thick green and magenta lines in Figure \ref{Fig:Energy100} in the bottom panel for $N=100$ particles. 
% \textcolor{purple}{Talk more about stationary points, hessian, characterization, degeneration, dof, classification, etc.} 
Some other representative separatrices are selected from \eqref{Separatrices} and displayed in the equation below. In particular, the separatrix $f_3$ provides a classification rule for the different chaotic regions (see Eqs.\eqref{DynamicalRegions},\eqref{SeparatricesLimit}) as we will see in the next sections. The other two separatrices presented in the equation below are relevant when studying the curvature of the density of states. 
%Limites de dominio calculados en 
%Stationary_points_LMG_U3.nb\Domain of the other separatrices
\begin{align}\label{Separatrix2}
	\begin{array}{llllr}
		f_3(\lambda)=\frac{(1- 3\lambda )^2}{10 \lambda },  && \forall \lambda \geq  \frac{1}{3} , && \mathrm{(ESQPT 3, blue)}\vspace{1mm}\\
        f_8(\lambda)=1-\lambda,  && \forall \lambda \geq  0, && \mathrm{(ESQPT 8, red)}\\
        f_9(\lambda)=1-\frac{3+\lambda^2}{6\lambda},  && \forall \lambda \geq  \frac{3}{4}. && \mathrm{(ESQPT 9, orange)}\\
%		f_4(\lambda)=\frac{(1+ \lambda )^2}{10 \lambda }, && \forall \lambda \geq \textcolor{red}{ \frac{3 }{5}} &&  \mathrm{(ESQPT 4)},\\
%		f_5(\lambda)=-1+\frac{1}{2\lambda}+\lambda, && \forall \lambda \geq \textcolor{red}{ \frac{1 }{2}} &&  \mathrm{(ESQPT 5)}.
	\end{array}
\end{align}
The ESQPT separatrices of Eqs.\eqref{Separatrix1} and \eqref{Separatrix2} are marked with representative colors (green, magenta, blue, red and orange in order) along the figures of this manuscript. See Appendices \ref{App:B}, \ref{App:C} for more details on the labeling and classification of the separatrices, about the domain where the separatrices are defined, and about the limitations imposed by the chosen patch of the complex projective manifold $\mathbb{C}P^2$ when defining the U(3) coherent states.

%%%%%%%%%%%%%%%%%%%%%%%%%%%%%%%%%%%%%%%5

\begin{figure}[h]
	\centering
	\includegraphics[width=0.48\textwidth]{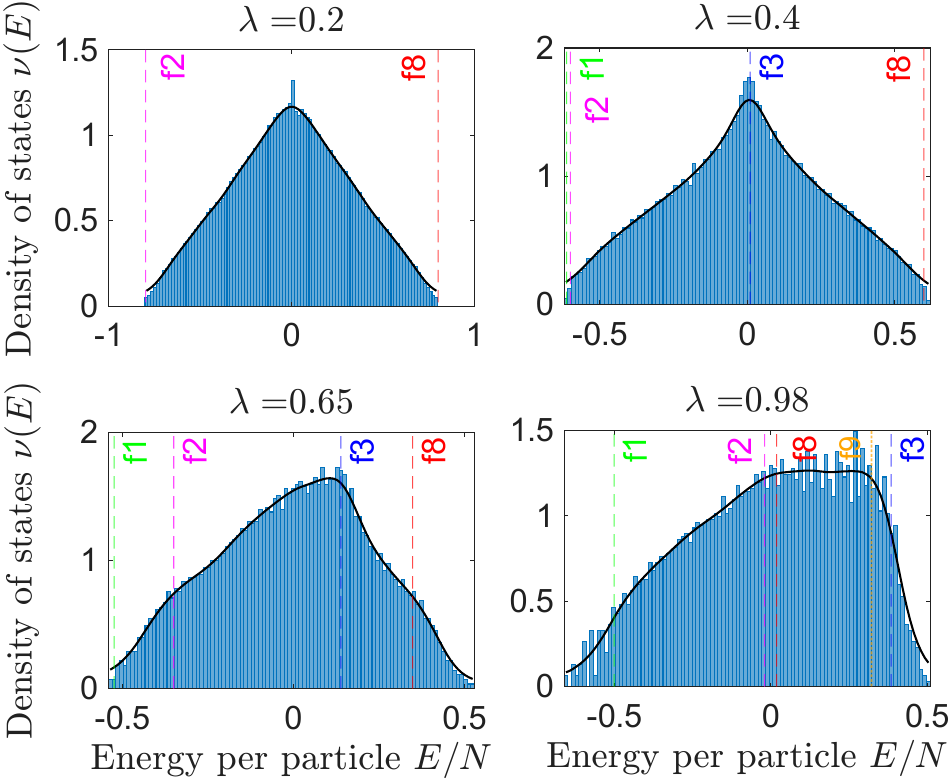}
    %Edit the pdf after exporting in Matlab: move the right column of figure to
%the left to reduce white space
	\caption{Histogram of 100 bins representing the density of states of the 3-level LMG model for $N=100$ particles and for some representative values of the control parameter $\lambda$ in the three QPT regions. The solid black line is the smoothed DOS with a MA window size of 15 elements. The dashed vertical lines represent the ESQPT separatrices \eqref{Separatrix1}, and \eqref{Separatrix2}.}
	\label{Fig:DOSmultiple100}
\end{figure}

%In the top panel of Figure \ref{Fig:Energy100}, the numerical energy densities for $N=15$ particles are plotted against the control parameter. We separate the four parity sectors in the diagonalization. There are regions in the center of the plot where the energies cluster creating the effect of ESQPT separatrices, but the complexity of the spectrum level crossings makes the classification difficult. The variational ground state energy \eqref{energysym} is traced in a dashed-dotted line, whose convergence with the numerical is enhanced in the bottom panel when the number of particles increases to $N=100$. Here we superimpose all separatrices \eqref{Separatrices} on the spectrum, adding colors to the most representative \eqref{Separatrix1} and \eqref{Separatrix2}. Below the ESQPT2 green line, there is a series of energies which degenerate when increasing $\lambda$ and create a recognizable pattern. 
In order to investigate and clarify the role of these ESQPT separatrices, the density of states (DOS) is computed from the energy spectrum of Figure \ref{Fig:Energy100} bottom panel, for selected fixed $\lambda$ values. The calculated DOS are depicted in Figure \ref{Fig:DOSmultiple100}. To construct this figure, we use 100 bin histograms and some representative values of the control parameter $\lambda=0.2, 0.4, 0.65,$ and $0.98$ in the three different QPT regions \eqref{energysym}, trying to avoid zones with an agglomeration of separatrices that can obscure the discussion. In Figure \ref{Fig:DOSmultiple100}, the colored vertical dashed lines are separatrices \eqref{Separatrix1} and \eqref{Separatrix2} marking changes in the shape and curvature of the DOS, often associated to ESQPT \cite{CejnarReview}. This is better represented in the smoothed DOS, which is the solid black line in the plot. We have used a Gaussian-weighted moving average (MA) filter to smooth the DOS, with a window size of 15 elements. This procedure will be detailed in the coming Section \ref{Sec:NNSD}. The deformation of the DOS when the control parameter is increased evinces how the interaction
destroys the symmetry of the Hamiltonian, $H$ \eqref{hamU3}. To better understand the role of the separatrices \eqref{Separatrix1} and \eqref{Separatrix2}, we study the chaoticity of the energy spectrum in the next section. 
% \textcolor{purple}{This is better represented in Figure \ref{Fig:DOSderivative100}, where the DOS is smoothed and we also plot its first derivative. In the classical limit, the ESQPT singularities can be characterized with the stationary points of the Hamiltonian \eqref{hamU3}. Two degrees of freedom, so the first derivative of the smoothed DOS detects the ESQPT and characterize the index of the ESQPT \cite{CejnarReview} (Pag31). The problem of this approach, is that we have to compute the index of all the stationary points vs lambda, and be careful if they are degenerate at some points. Furthermore, the derivative of the smoothed DOS does not detect very well the ESQPT separatrices, check out Figure \ref{Fig:DOSderivative100}. Hence, I recommend to skip this discussion of the derivative as in the anharmonic Lipkin article. Ver} \url{https://www.dropbox.com/home/3-level%20Lipkin/MatLab}.

% \begin{figure}[h]
% 	\centering
% 	\includegraphics[width=0.5\textwidth]{Figures/DOS_multipleLambda100_0_0_lambda_0.001_1_0.01.pdf}
% 	\caption{Density of states of the 3-level LMG model for $N=100$ particles and different values of the control parameter $\lambda$. The dashed vertical lines represent the ESQPT separatrices \eqref{Separatrix1},\eqref{Separatrix2}.\textcolor{blue}{Ver si merece la pena estudiar caos y $\eta$ en $\lambda=0.65$, o si no es posible como dice Meredith porque no hay secciones de Poincare con trayectorias caoticas aisladas sin orbitas integrables.}}
% 	\label{Fig:DOSmultiple100}
% \end{figure}

% Not to use widetext environment
% https://tex.stackexchange.com/questions/156590/how-can-i-make-my-caption-span-both-columns-in-a-twocolumn-revtex-document
% \begin{figure*}
% 	\centering
% 	\includegraphics[width=1\textwidth]{Figures/DOS_derivative_yyaxis_100_0_0_numBins200_MA15lambda_0.001_1.02_0.01.pdf}
% 	\caption{Smoothed density of states of the 3-level LMG model for $N=100$ particles and different values of the control parameter $\lambda$. We use Moving Average 15 for smoothing over a DOS histogram of 200 bins. The dashed vertical lines represent the ESQPT separatrices \eqref{Separatrix1},\eqref{Separatrix2}.}
% 	\label{Fig:DOSderivative100}
% \end{figure*}
%\end{widetext}

%%%%%%%%%%%%%%%%%%%%%%%%%%%%%%%%%%%%%%%%%%%%%%%%%%%%%%%%%%%%%%%%%%%%%%%%%%%%%%%%%%%%%%%%%%%%%%%%%

%\textcolor{purple}{Pepe - Voy por aquí}

\section{Results}\label{sec4}

\subsection{Poincar\'e sections and Peres lattices}
If a quantum system is regular or non-chaotic, then its classical analog system is integrable. Therefore, to study the regularity of the 3-level LMG Hamiltonian, one should examine the dynamics of the model in its classical limit. We can arrive at the classical LMG Hamiltonian in terms of its canonical momentum and position variables, $p$ and $q$ respectively, by making use of the following changes of variables:
\begin{equation}\label{changeCoor1}
   \beta_j \equiv \frac{z_j}{(1+|z_1|^2 + |z_2|^2)^{1/2}}, ~j= 1,2,
\end{equation}
and then
\begin{equation}\label{changeCoor2}
    \beta_j = \frac{q_j + ip_j}{\sqrt{2}}.
\end{equation}

Applying these to Eq.~(\ref{enersym}), we arrive at
\begin{align}
    E(q,p,\lambda) = -&(1 - \lambda) + \scriptstyle\frac{1}{2} \textstyle q_1^2(1 - 3 \lambda) \nonumber \\
    +& \scriptstyle\frac{1}{2} \textstyle q_2^2 (2 -4 \lambda) \nonumber 
    + \scriptstyle\frac{1}{2} \textstyle p_1^2 (1 + \lambda) + \textstyle p_2^2 \nonumber \\
    +& \scriptstyle\frac{1}{2} \textstyle\lambda \left((q_1^2 + q_2^2)^2 - (p_1^2 + p_2^2)^2 \right. \nonumber \\
    -& \left.(q_1^2 -  p_1^2) (q_2^2 - p_2^2) - 4 q_1 q_2 p_1 p_2 \right).
       \label{eq:classical_H}
\end{align}

Now, if a system with two degrees of freedom is integrable, then there is a second conserved quantity in addition to the energy, which forces the trajectories to lie on a two-dimensional torus in the phase space. If we take a Poincaré surface section of these trajectories, in this case, the intersection of the trajectories with a given plane in the
phase space (e.g. $q_1=q_{10}$) for a particular $\lambda$ and energy value,
%the intersection of the trajectories with a plane of a particular $\lambda$ and energy value,
these tori manifest as closed curves within the plane. Meanwhile, if the system is chaotic, the Poincaré section consists of a set of points evenly covering the available phase space without any particular structure. For any case that lies in between, the sections will have chaotic regions of evenly distributed points as well as closed curves.

\begin{figure}
	\centering
	\includegraphics[width=0.5\textwidth]{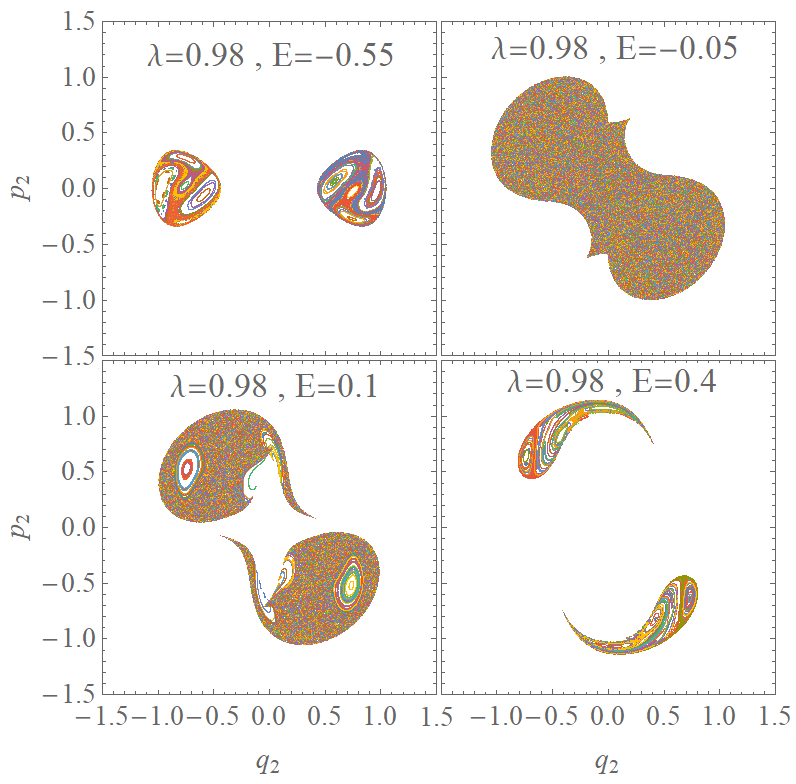}
	\caption{Poincar\'e sections of the energy surface in the position and momentum space \eqref{eq:classical_H}, for a fixed value of the control parameter $\lambda=0.98$ and some representative values of the energy $E$ inside each dynamical region \eqref{DynamicalRegions}. The sections are plotted in the $(q_2, p_2)$ space for crossings with the $q_1=0.8165$ plane, the equilibrium value of $q_1$. Quasi-integrable trajectories are depicted in the diagonal subplots of the grid. Quasichaotic and chaotic trajectories are shown in the bottom left and top right subplots respectively. }
	\label{Fig:PoincareSections}
\end{figure}

In Figure~\ref{Fig:PoincareSections} we represent the $(q_2,p_2)$ Poincaré sections of the classical Hamiltonian in Eq.(\ref{eq:classical_H}) at $\lambda = 0.98$. The sections cut the plane $q_1=q_{10}$, where the equilibrium value $q_{10}^2=2/3$ is obtained from the potential energy surface $E(q,p=0,\lambda)$ minimization. Furthermore, the conservation of the total number of particles $q_1^2+q_2^2+p_1^2+p_2^2\leq 2$ and $\dot{q}_1\geq 0$ are necessary constraints for the Poincaré section calculation \cite{Meredith}.

The energies chosen in Figure~\ref{Fig:PoincareSections} lie in the regions between the ESQPTs defined in Eqs.~(\ref{Separatrix1}),~(\ref{Separatrix2}) and correspond to chaotic (top right panel), quasi-chaotic (bottom left) and quasi-integrable (diagonal panels) regions. That is, we propose an analytical model for $\lambda\simeq 1$ which demarcates the dynamical classes using ESQPT separatrices, highlighting the connection between chaos and ESQPT in some specific cases \cite{PhysRevE.83.046208,PhysRevE.94.012140,CAPRIO20081106,PhysRevE.78.031130,PhysRevA.111.L031502},

\begin{align}
	\begin{array}{llllr}
		\text{Quasi-int.} \quad&& E\leq -\frac{1}{2}  && (E_0\:\mathrm{ to }\: f_1),\\[1ex]
		\text{Chaotic}\quad&&  E\in(-\frac{1}{2},0]  && (f_1\:\mathrm{ to }\: f_2),\\[1ex]
		\text{Quasicha.} \quad&& E\in(0,\frac{1}{4}]  && (f_2\:\mathrm{ to } \: f_3),\\[1ex]
		\text{Quasi-int.}\quad && E> \frac{1}{4} && (f_3\: \mathrm{ to }\: E_{\textrm{max}}).\\
	\end{array}\label{DynamicalRegions}
\end{align}
% \begin{align}
% 	\begin{array}{llllr}
% 		\text{Quasi-int.} \quad&& E\leq-\frac{(1+ \lambda )^2}{8 \lambda }\simeq -\frac{1}{2}  && (E_0\:\mathrm{ to }\: f_1),\\
% 		\text{Chaotic}\quad&&  E\in(-\frac{(1+ \lambda )^2}{8 \lambda },-1+\lambda]\simeq(-\frac{1}{2},0]  && (f_1\:\mathrm{ to }\: f_2),\\
% 		\text{Quasicha.} \quad&& E\in(-1+\lambda,\frac{(1- 3\lambda )^2}{10 \lambda }]\simeq (0,\frac{1}{4}]  && (f_2\:\mathrm{ to } \: f_3),\\
% 		\text{Quasi-int.}\quad && E>\frac{(1- 3\lambda )^2}{10 \lambda }\simeq \frac{1}{4} && (f_3\: \mathrm{ to }\: E_{\textrm{max}}).\\
% 	\end{array}\label{DynamicalRegions}
% \end{align}

% \textcolor{blue}{Proposal: Make the same classification for other values of lambda, previously analyze the Poincare section and Peres lattices to delimit the regions. Use the last paragraph of left column page 5 of Meredith.} 
This classification for $\lambda\simeq 1$ is shown in the bottom panel of Figure \ref{Fig:Energy100}, where different groups of separatrices merge with the colored ones in $\lambda\to 1$, simplifying the spectrum classification in dynamical regions (see Eq.\eqref{SeparatricesLimit} and Appendix \ref{App:B} for a detailed discussion). It is important to note the necessity of selecting a value of the control parameter $\lambda=0.98\simeq 1$, where we have a reasonable energy range with quasi-integrable and completely chaotic dynamics separately \cite{Meredith}. We include a collection of animations of the Poincaré section plots for different values of $\lambda$ and the energy in the supplemental material~\cite{zenodo_Poincare_sec_videos}, where the lack of a wide energy range of integrable states can be observed. 
%(see \cite{Meredith} for more details)

The limitation to $\lambda\simeq 1$ is also required in Figures \ref{Fig:NNSDunfolded}, \ref{Fig:eta_unfolded_p_eee}, \ref{Fig:PS100}, and \ref{Fig:Overlap100} to display regions with a reasonable chaotic or quasi-integrable character. In fact, the selected $\lambda=0.98$ coincides with the control parameter $\chi=100$ chosen in the bibliography \cite{lipkin1,lipkin2,Meredith,KusLipkin}, where the reparameterization $\lambda=\frac{\chi/2}{1+\chi/2}$ is necessary to map the Hamiltonian \eqref{hamU3old} to our model \eqref{hamU3}. We also mention that the role of the separatrix $f_8$ \eqref{Separatrix2} in the classification \eqref{DynamicalRegions} is neglected as $f_2(\lambda)\simeq f_8(\lambda)$ for the case of study $\lambda=0.98$ (see Appendix \ref{App:B} for more details.). The distinction between chaotic and quasichaotic dynamics is described quantitatively for $\lambda=0.98$ in \cite{Meredith}, using the chaotic volume and the Lyapunov exponent as measures for the semiclassical approximation of the 3-level LMG model. More explicitly, this criterion associates the intervals $(0,0.7),(0.7,0.95),(0.95,1)$ of the chaotic volume with the quasi-integrable, quasichaotic and chaotic regions respectively. We use this prescription in equation \eqref{DynamicalRegions} and along this work. It can be checked that the energies proposed in \cite{Meredith} to classify the dynamical regions are equivalent to our proposal in eq.\eqref{DynamicalRegions}, where the mapping $E(\lambda)=(1-\lambda)\tilde{E}(\tfrac{2\lambda}{1-\lambda})$ is applied to the reference energies $\tilde{E}$. Nevertheless, we will check if this classification is consistent with other measures of chaos in the next sections.

The criterion in \eqref{DynamicalRegions} is also supported by the Morse index of the separatrices computed in Table \ref{tab:r} (see App.\ref{App:B} for more details). In fact, we obtain the positive result that the Morse index correlates with relevance to chaos at $\lambda\simeq 1$. That is, for the three selected separatrices $f_1,f_2,f_3$ in the dynamical region criterion \eqref{DynamicalRegions}, the Morse index is $r=1, 2, 3$ respectively in the region $\lambda\in[3/4,1]$. Hence, the changes in the Morse index of the separatrices indicate the transition between dynamical regions. This is summarized in Table \ref{tab:MorseAndChaos}, including the groups of separatrices that converge at $\lambda\to 1$ (see also eq.\eqref{SeparatricesLimit}). Note that, despite the limit $\lim_{\lambda\to 1}f_9(\lambda) = 1/3$ does not coincide with any of the energy clusters defining the dynamical regions (-2/3, -1/2, 0, 4/10, 1/2), it becomes irrelevant in the classification, since its Morse index $r=2$ is the same as the indices of $f_2,f_7,f_8$.

\begin{table}[H]
%Retrieved from Stationary_points_LMG_U3_v2.nb
$$
\begin{array}{|c|c|c|c|}
\hline
&
r_i &
f_i(\lambda\simeq 1) &
\text{Similar} 
\\
\hline

f_0 &
0 &
-\frac{2}{3} &
\\[0.5ex]
\hline

f_1 &
{\color{green}1} &
-\frac{1}{2} &
f_5,f_6
\\[0.5ex]
\hline

f_2 &
{\color{magenta}2} &
0 &
f_7,f_8
\\[0.5ex]
\hline

f_3 &
{\color{blue}3} &
\frac{4}{10} &
f_{10},f_{11}
\\[0.5ex]
\hline

f_4 &
\textcolor{darkbrown}{4} &
\frac{1}{2} &
f_{12},f_{13}
\\[0.5ex]
\hline

\end{array}
$$
    \caption{From left column to right, Morse indices $r_j$ (extracted from Table \ref{tab:r}), separatrices energy (see Eq.\eqref{SeparatricesLimit}), and similar separatrices to the main separatrices selected in the dynamical region criterion \eqref{DynamicalRegions} for $\lambda\to 1$. Each color represents a different value of the indexes. The table displays how Morse index correlates to chaos for the limit case $\lambda\to 1$.}
    \label{tab:MorseAndChaos}
\end{table}

In addition to the Poincaré sections, we also study some of the so-called Peres lattices of the system. When the classical analog $E(q,p,\lambda)$ is integrable, the Einstein–Brillouin–Keller (EBK) quantization becomes a good approximation in the semiclassical limit, making it so that the energy spectra is regularly distributed. This can easily be seen in Fig.~\ref{Fig:Energy100} for $\lambda=0$, where the system is integrable, and the energies are equally spaced, but becomes unclear as $\lambda$ increases, and the system presents more chaotic regions. Nevertheless, this regularity property holds for the eigenvalues of any conserved dynamical variable, not only the Hamiltonian, and therefore, by plotting the expected values of these conserved quantities vs the energy spectra, the chaos of the system can be studied. If the system is integrable, then the EBK approximation holds and both observables will be distributed in a regular lattice when plotted against each other, while failure to do so is indicative of quantum chaos~\cite{Peres1984}.

\begin{figure}
	\centering
	\includegraphics[width=0.48\textwidth]{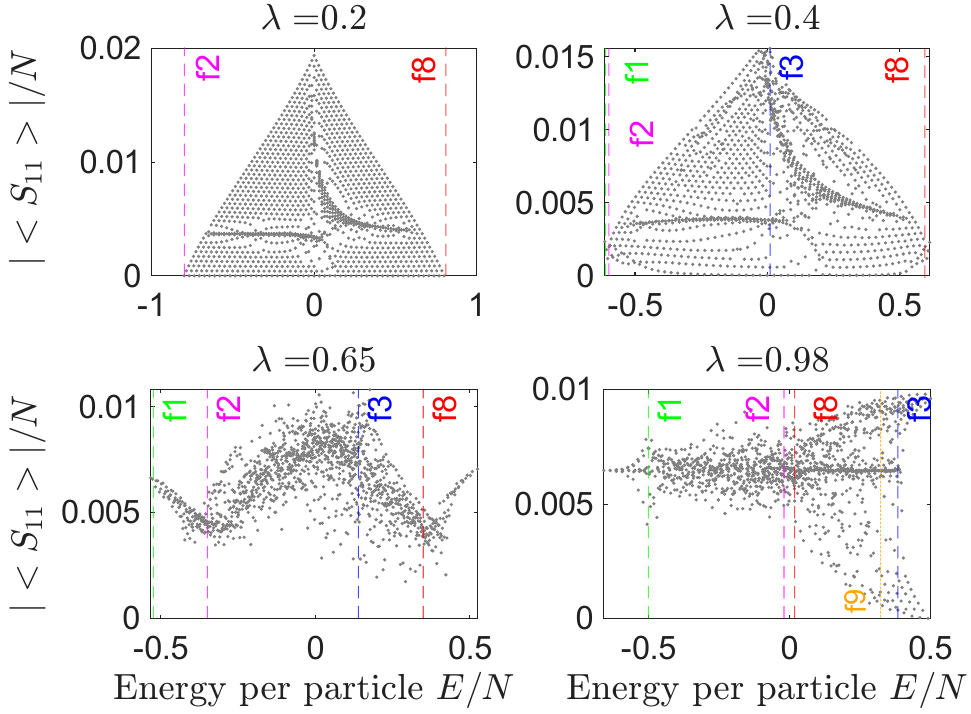}
	\caption{Peres lattices for the expected values $|\la S_{11}\ra/N|$ vs the different excitation energies of the 3-level LMG model. We choose $N=100$ particles, some representative values of the control parameter $\lambda$ in each column, and we restrict to the parity sector $p=(+,+,+)$ for the sake of clarity. The dashed vertical lines represent the ESQPT separatrices \eqref{Separatrix1}, \eqref{Separatrix2}. Similar results are obtained for $\la S_{22}\ra$, $\la S_{00}\ra$ and $\la S_{22}-S_{00}\ra$.}
	\label{Fig:PeresLattices100}
\end{figure}

Here we consider the expected values of the collective spin operators $S_{ii}$. In particular, Fig.~\ref{Fig:PeresLattices100} shows the Peres lattices of $|\la S_{11}\ra/N|$ vs the energy per particle of the Hamiltonian. Additionally, ESQPT 1, 2, 3, 8 and 9 separatrices \eqref{Separatrix1}, and \eqref{Separatrix2} are also represented. We have already seen from the energy spectra that the system is integrable for $\lambda = 0$ and that, at higher values of $\lambda$, chaotic and quasi-chaotic regions start to appear, but notably, Fig.~\ref{Fig:PeresLattices100} shows that quasi-chaotic regions already appear at small values of $\lambda$, which was not clearly seen through the density of states. The results for other $S_{ii}$ operators show similar trends.

Finally, the chosen separatrices can be used to delimit the different regions at high values of $\lambda$ as we move up from the ground state to higher excited states, being quasi-integrable up to ESQPT1, then chaotic from ESPQT1 to ESPQT2, quasi-chaotic from ESPQT2 to ESPQT3, and again quasi-integrable after ESQPT3. As expected, for middle values of $\lambda$, delimiting these regions is much harder as the spectrum is more complex, ESQPTs cross each other, and the regions overlap. Nevertheless, one can still qualitatively identify some chaotic, quasi-chaotic, or quasi-integrable regions from the distribution of points in the Peres lattices.

\subsection{Unfolded spectrum and nearest-neighbors spacing distribution}\label{Sec:NNSD}

%chapter 5.19 of Haake
We perform an unfolding procedure to smooth the DOS and extract a local average function of it \cite{Haake}. It allows us to erase spectral fluctuations and compare spectra in an unbiased manner \cite{Meredith}. Firstly, we compute a histogram of the DOS $\rho(E)$ (using numerical diagonalization data) and smooth it $\rho_{\Delta}(E)$ with a Gaussian filter of variance $\Delta^2$. In practice, we set the window size of a Gaussian-weighted moving average (MA) filter in place of $\Delta^2$. Then we calculate the cumulative distribution function of the numerical data $\sigma(E)$ (level staircase), and the integral of the smoothed DOS $\sigma_{\Delta}(E)$ (trapezoidal numerical integration). After that, we compare both distributions $\sigma(E)$ and $\sigma_{\Delta}(E)$ using a two-sample Kolmogorov-Smirnov test \cite{Stephens01091974}, and choosing the optimal window size that minimizes the test error. Eventually, we fit the values of $\sigma_{\Delta}(E)$ with a cubic spline to obtain a function defined in a continuous domain $E\in\mathbb{R}$, and identify the unfolded spectrum as $e_i=d\cdot \sigma_{\Delta}(E_i)$, with $d$ the Hilbert space dimension and $E_i$ the numerical energies.

We use a MA window size of 15 elements and $N=100$ to calculate the smoothed DOS $\rho_{\Delta}(E)$, over a histogram of 200 bins for the DOS $\rho(E)$. Then, we perform the Kolmogorov-Smirnov test that gives a statistic equal to 0.015 and a $p$-value equal to 1, so we fail to reject the null hypothesis that the two samples $\sigma(E)$ and $\sigma_{\Delta}(E)$ are from the same continuous distribution at the 5\% significance level.
%Check Matlab, SCRIPT_LMG_U3_diagonalization_EE_parity_basis_v2.m\kstest2

% \textcolor{purple}{In Figure \ref{Fig:DOSderivative100}, the solid black lines represent the smoothed DOS with a MA window size of 15 elements and $N=100$. Note that the smoothing has been performed over a histogram of 200 bins, rather than the 50 bins used in Figure \ref{Fig:DOSmultiple100} for the sake of clarity. Compute the derivative of spline smoothed DOS and add here some comments. However, I think that the derivative is not going to give us any qualitative identification of ESQPTs.} 

\begin{figure}
	\centering
	\includegraphics[width=0.48\textwidth]{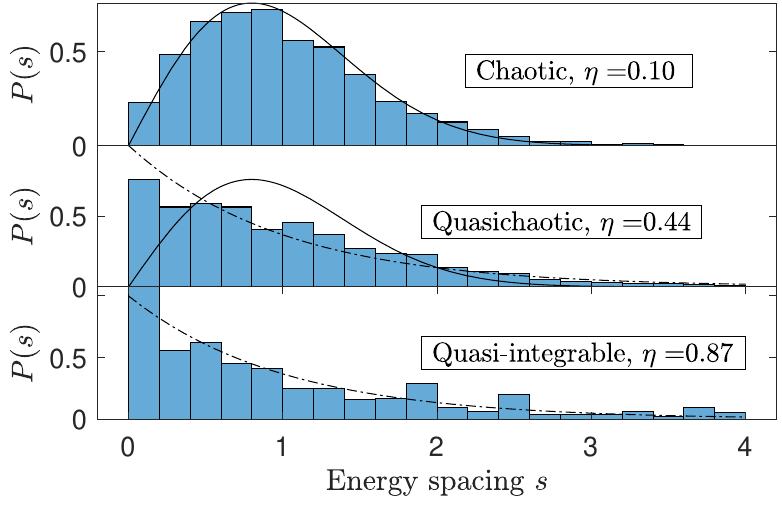}
	\caption{Nearest-neighbors spacing distribution of the 3-level LMG model for $N=100$ particles, and $\lambda=0.98$. The unfolded spectrum is computed over a DOS histogram of 200 bins, smoothed by a Gaussian filter with a moving average window of 15 elements \cite{Haake}. The energies included in the NNSD calculation of each subplot are delimited by the regions defined in \eqref{DynamicalRegions}. That is, the bottom panel represents the states in quasi-integrable regions $(E_0,f_1)$ and $(f_3,E_\text{max})$, the top panel in the chaotic $(f_1,f_2)$ region, and the middle panel in the quasichaotic $(f_2,f_3)$ region. We also display the $\eta$ degree of chaos \eqref{eta} on top, and plot the Wigner (solid) and Poisson (dashed-dot) distributions.}
	\label{Fig:NNSDunfolded}
\end{figure}

Using the normalized energy spacings $s_i=(e_{i+1}-e_i)/\bar{e}$ of the unfolded spectrum, we can define the nearest-neighbor spacing distribution (NNSD) as
the probability $P(s)$ that two neighboring eigenvalues
are a distance $s$ apart. The energy levels are classified into the dynamical classes defined in equation \eqref{DynamicalRegions}: chaotic, quasichaotic, and quasi-integrable. The chaotic ones will have a NNSD with Wigner distribution shape $P_W(s)=\frac{\pi}{2}s\exp(-\tfrac{\pi}{4}s^2)$, and the quasi-integrable ones a Poisson distribution shape $P_P(s)=\exp(-s)$. This is presented in Figure \ref{Fig:NNSDunfolded} for $N=100$ and $\lambda=0.98$, where the three dynamical classes are shown in each subplot, and the solid and dashed-dotted lines represent the Wigner and Poisson distributions respectively. The pictures also present a qualitative measure of the degree of chaos, the parameter $\eta$, which is introduced in the coming paragraph \eqref{eta}. To obtain a better fit of the histograms to the distributions, we use 200 bins in the DOS and a MA window of 15 elements when smoothing. In addition, the level classification \eqref{DynamicalRegions}, the unfolded spectrum and the energy spacings are conveniently computed in each parity sector of the Hamiltonian, and then the results are merged to calculate the NNSD histogram of Figure \ref{Fig:NNSDunfolded}. The bottom panel represents the states in quasi-integrable regions $(E_0,f_1)$ and $(f_3,E_\text{max})$, the top panel in the chaotic $(f_1,f_2)$, and the middle panel in the quasichaotic $(f_2,f_3)$.
%\textcolor{blue}{Talk about Random Matrix Theory, Gaussian Orthogonal Ensamble vs Poisson statistics, read the Intro of \cite{Meredith}?} 

% Eliminar a mano los 1's del eje y, pinchar en el cuadro de texto del pdf y borrar

\begin{figure}
	\centering
	\includegraphics[width=0.48\textwidth]{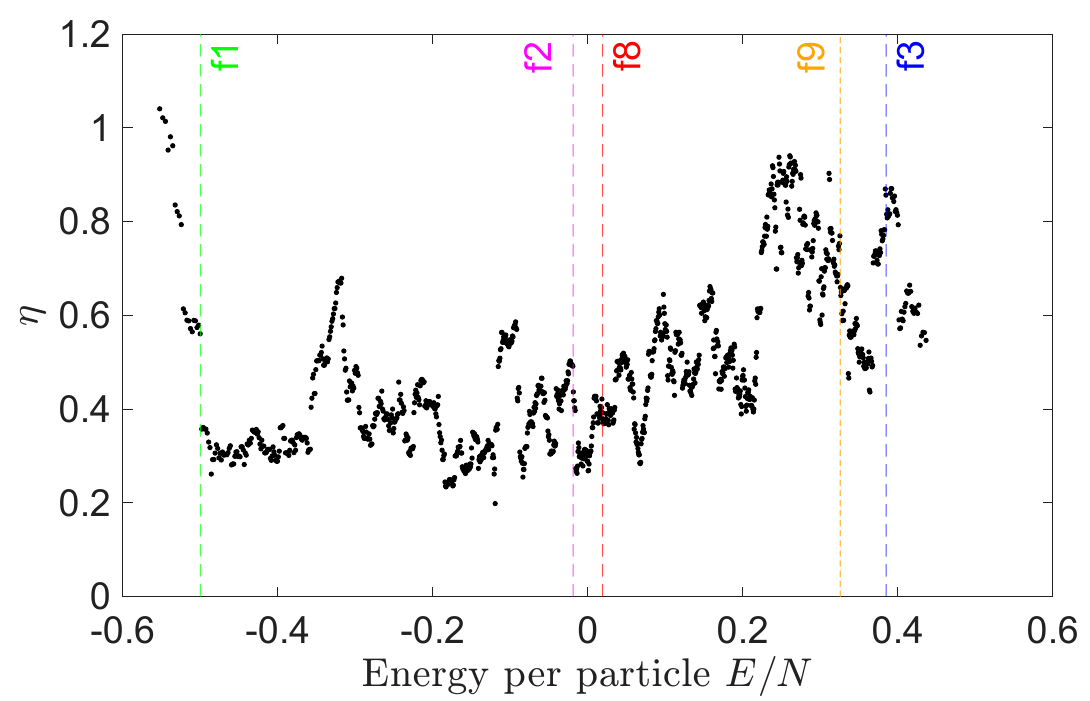}
	\caption{Degree of chaos $\eta$ \eqref{eta} of the 3-level LMG model for $N=100$ particles, $\lambda=0.98$, and parity $p=(+,+,+)$. To compute each value of $\eta$, we select sequences of 70 energy spacings and calculate the corresponding NNSD. Previously, the spectrum is unfolded over a DOS histogram of 200 bins, smoothed by a Gaussian filter with a 15 elements window \cite{Haake}. The dashed vertical lines represent the ESQPT separatrices \eqref{Separatrix1}, and \eqref{Separatrix2}.}
	\label{Fig:eta_unfolded_p_eee}
\end{figure}

In order to quantitatively delimit the energy range of each class, we measure the degree of chaos with the following expression \cite{PhysRevC.95.054326}
\begin{equation}
	\eta=\frac{\sigma_s-\sigma_W}{\sigma_P-\sigma_W},
	\label{eta}
\end{equation}
with $\sigma_W=4/\pi-1$ and $\sigma_P=1$ the variance of the Wigner and Poisson distribution respectively, and $\sigma_s$ the variance of the NNSD in question. More explicitly, we select several consecutive spacings $s$ (200 for example, without considering the dynamical regions \eqref{DynamicalRegions}), compute the NNSD and its associated value of $\eta$ \eqref{eta}, and repeat the process for a new sequence of spacings. The values $\eta$ are plotted versus the mean energy of each sequence in Figure \ref{Fig:eta_unfolded_p_eee}, for the parity $p=(+,+,+)$, $N=100$ particles and $\lambda=0.98$. While the analysis of individual parity sectors yields similar results, the superposition of all parity sectors obscures the characterization of the system’s chaoticity. In this figure, we compute the unfolded spectrum for a DOS histogram with 200 bins, a MA window of 15 elements, and to calculate $\eta$ we select sequences of 70 energy spacings. High values of $\eta$ appear to the left and right of the separatrices $f_1$ (green) and $f_3$ (blue) respectively, corresponding to the integrable regime (Poisson distribution $\sigma_P=1$). In contrast, the degree of chaos has reduced values for the energies between these two separatrices, which is characteristic of a chaotic regime (Wigner distribution $\sigma_W\simeq 0.27$). 

\begin{figure}
	\centering
	\includegraphics[width=0.5\textwidth]{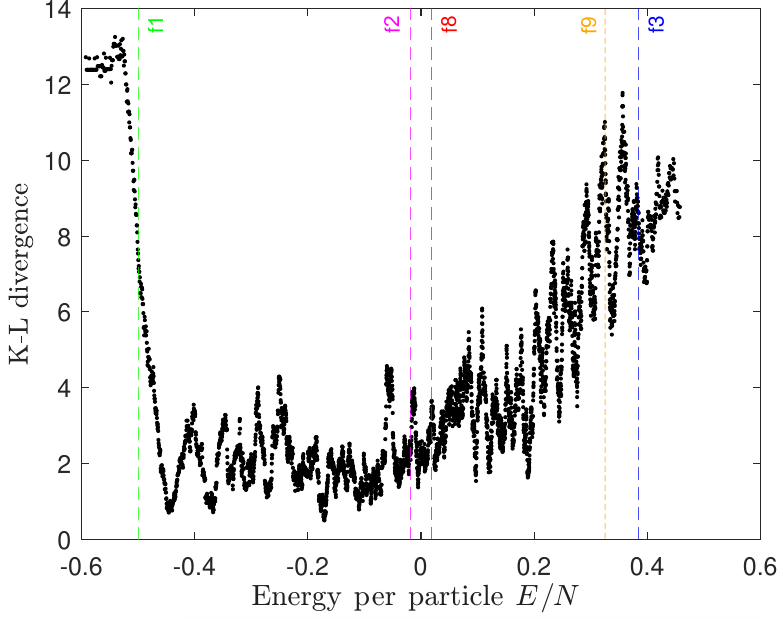}
    \includegraphics[width=0.5\textwidth]{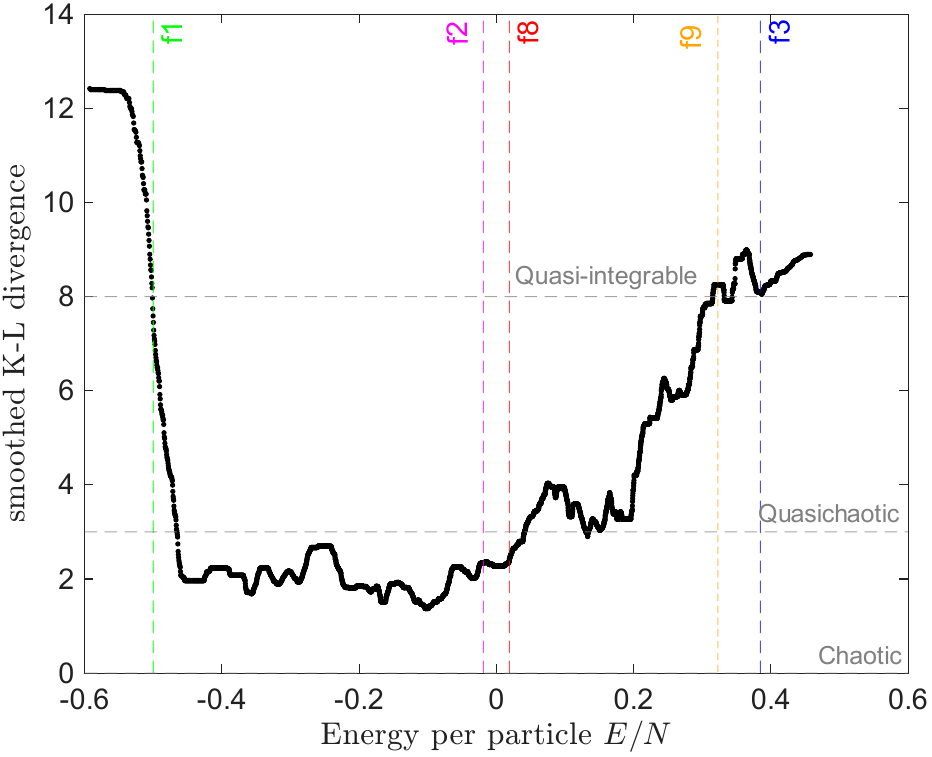}
	\caption{ (Top) Kullback–Leibler divergence of the 3-level LMG model for $N=100$ particles and $\lambda=0.98$. To compute each value of $D_{\text{KL}}$, we select sequences of 70 energy spacings and calculate the corresponding NNSD (20 bins histogram with $s_{\text{max}}=4$). Previously, the spectrum was unfolded over a DOS histogram of 200 bins, smoothed by a Gaussian filter with a 15 elements window \cite{Haake}. The dashed vertical lines represent the ESQPT separatrices \eqref{Separatrix1}, and \eqref{Separatrix2}. (Bottom) Smoothed version of the KL divergence of the top figure, using a moving median with a 200 elements window. The horizontal grid lines delimit the dynamical regions of \eqref{DynamicalRegions}. }
	\label{Fig:KLdivergence_unfolded}
\end{figure}

We also include a second measure of chaoticity, called the Kullback–Leibler (KL) divergence \cite{Kullback-Liebler}
\begin{equation}
    D_{\text{KL}}(P||Q)=\sum_s P(s)\log(\tfrac{P(s)}{Q(s)}),
\end{equation}
where $P(s)$ is the NNSD and $Q(s)=P_W(s)$ is the Wigner distribution. The KL divergence compares the NNSD with a Wigner distribution; that is, it gives a value of $0$ for chaotic distributions and values greater than zero otherwise. On the top panel of Figure \ref{Fig:KLdivergence_unfolded}, the KL divergence is plotted for the same parameters as those used in the degree of chaos Figure \ref{Fig:eta_unfolded_p_eee}, but using all parity sectors at once. We observe a drop in the KL divergence when crossing the first separatrix $f_1$ (green), delimiting quasi-integrable and chaotic regimes. Then, after the separatrices $f_2$ (magenta) and $f_8$ (red), the divergence starts growing again in the quasichaotic regime, and stabilizes after the separatrix $f_3$ (blue) in the second quasi-integrable regime zone. The KL divergence gives better qualitative evidence of chaoticity and ESQPT than the degree of chaos. Furthermore, we plot on the bottom panel of Figure \ref{Fig:KLdivergence_unfolded} a smoothed version of the KL divergence with the same parameters, using a moving median smoothing of 200 elements per window. With the horizontal grid lines in the plot, we set an arbitrary bound of the KL divergence to separate dynamical regions according to the equation \eqref{DynamicalRegions}. This serves as a quantitative procedure (for finite $N$) to classify which energy levels have statistics compatible with random matrix statistics, that is, a Wigner shape NNSD. However, these boundaries vary with the chosen number of particles $N$, hence a general procedure requires the usage of the semiclassical model, the chaotic volume and the Lyapunov exponent \cite{Meredith}.

The unfolded spectrum is also used in Appendix \ref{App:B} to give insight into the ESQPT classification according to the $r-$index and the smoothed DOS first derivative.

\subsection{Participation ratio}
\begin{figure}[h]
	\centering
	\includegraphics[width=0.5\textwidth]{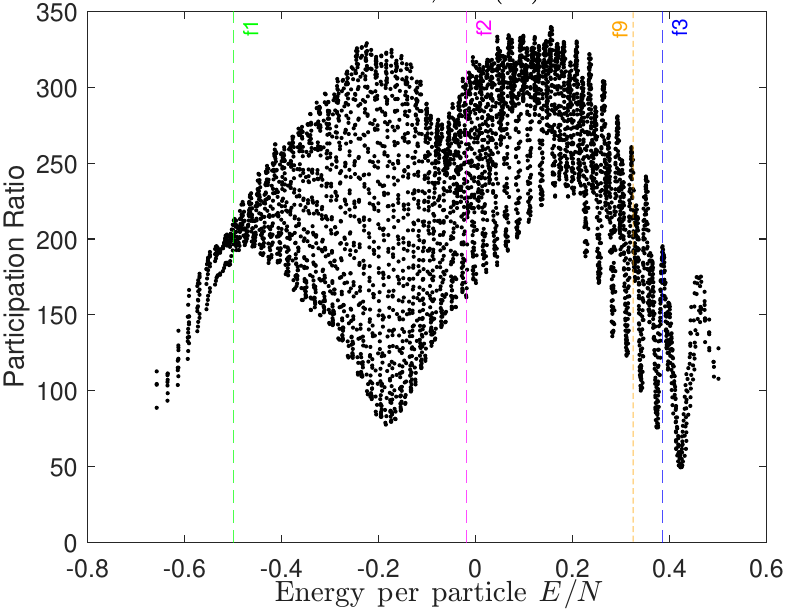}
	\caption{Participation ratio of the 3-level LMG model  for $N=100$ and $\lambda=0.98$. Each dot represents the PR of a numerical eigenstate with energy $E$. The plot is smoothed with a MA of 15 elements window. The dashed vertical lines represent the ESQPT separatrices \eqref{Separatrix1}, and \eqref{Separatrix2}.}
	\label{Fig:PS100}
\end{figure}

We also characterize the ESQPT with the Participation Ratio (PR), which is defined as 
$$
PR(\lambda,E_j)=\frac{1}{\sum_{\|\vec{n}\|_1=N}\,|c_{\vec{n},j}(\lambda)|^4},
$$
with $c_{\vec{n},j}$ the components of the eigenstates $|\psi_j(\lambda)\ra$ expressed in the Fock basis,  with energies $E_j$. It is plotted for $N=100$ and $\lambda=0.98$ in Figure \ref{Fig:PS100}, where we perform a MA smoothing with 15 elements window to damp the strong oscillations of the PR. Following the classification in \eqref{DynamicalRegions}, there is a PR cluster to the left of $f_1$ corresponding to the quasi-integrable states. The PR points spread to the right of this separatrix, introducing the chaotic region. The line $f_2$  is close to a local minimum in the top points with higher PR, characterizing the ESQPT2 and the beginning of the quasichaotic region. The last line $f_3$ identifies a local drop of the PR corresponding to the ESQPT3 and the start of the second quasi-integrable region.

\subsection{Overlap distribution}

\begin{figure}[h]
	\centering
	\includegraphics[width=0.5\textwidth]{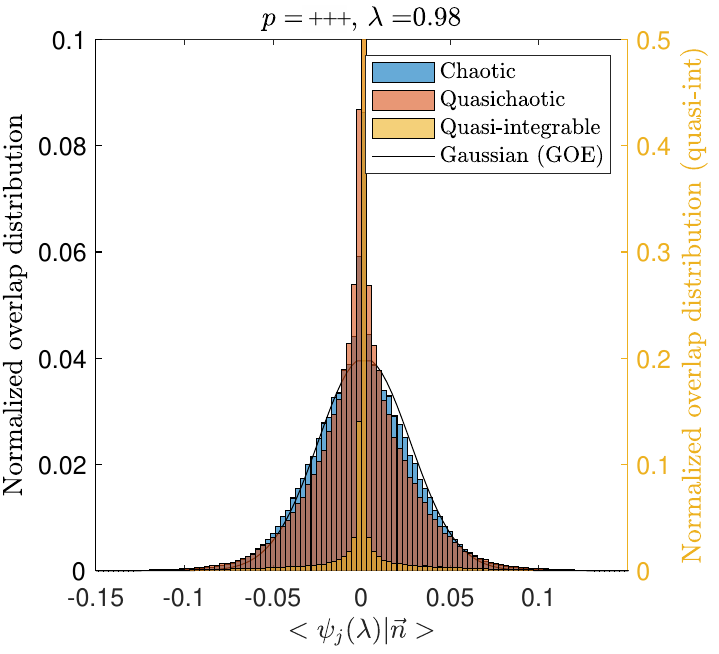}
	\caption{Normalized overlap distribution $\la \psi_j(\lambda)|\vec{n}\ra$ of the 3-level LMG model for $N=100$, $\lambda=0.98$ and the parity sector $p=(+,+,+)$. The different colors represent the dynamical regions in \eqref{DynamicalRegions}, with the left y-axis referring to chaotic and quasichaotic regimes, and right y-axis to quasi-integrable. Similar results are obtained for the other parity sectors \eqref{parityEven}.}
	\label{Fig:Overlap100}
\end{figure}

The overlap distribution is computed as the scalar product of an eigenstate of the LMG Hamiltonian $|\psi_j(\lambda)\ra$ \cite{Meredith}, $j=1,\ldots,d$, and the basis states $|\vec{n}\ra$ of the Hilbert space \eqref{symmetricbasis}. The overlap $x_{j,\vec{n}}(\lambda)=\la\psi_j(\lambda)|\vec{n}\ra$ is computed for a fixed value of the control parameter $\lambda$ and for all the combinations of the indexes $j$ and $\vec{n}$, giving a total of $d^2$ values, whose frequencies are represented with a normalized distribution in the histogram of Figure \ref{Fig:Overlap100}. With the yellow color, we represent states in the quasi-integrable regime for $N=100$, $\lambda=0.98$ and parity $p=(+,+,+)$, which have a delta-like distribution since
an integrable system is expressed in terms of only a few states of the basis. In blue and red, chaotic and quasichaotic states resemble a Gaussian shape, which is characteristic of the states in a Gaussian Orthogonal Ensemble (GOE), so the overlap $x_{j,\vec{n}}$ can be seen as a Gaussian random variable in this case \cite{PhysRevLett.111.037001,Haake}. Similar results are obtained for the rest of the parities \eqref{parityEven}.

% https://journals.aps.org/prl/abstract/10.1103/PhysRevLett.111.037001
%Here we announce and explain an unexpected discovery: sequences of fermion-parity switches are not independent. As illustrated in Fig. 1, for a quantum-dot model Hamiltonian [16], the level crossings show an antibunching effect, with a spacing distribution that vanishes at small spacings. This is reminiscent of level repulsion, but we find that the spacing distribution is distinct from the Wigner distribution of random Hamiltonians [17,18]. Instead, it is a hybrid of the Wigner distribution (linear repulsion) for small spacings and the Poisson distribution (exponential tail) for large spacings. This Wigner-Poisson statistics has appeared once before in condensed matter physics at the Anderson metal-insulator transition [19,20]. We construct an ensemble of non-Hermitian matrices that describes the hybrid statistics, in excellent agreement with simulations of a microscopic model.

% http://dx.doi.org/10.1103/PhysRevLett.111.090404
%While in chaotic
%quantum systems the spectrum features level repulsion,
%regular behavior is related to level clustering. These features
%are nicely displayed by the unfolded distribution [17]
%of the level spacings s in the spectrum. A Poisson
%distribution PðsÞ ¼ es indicates level clustering, while
%chaotic Hamiltonians with time-reversal invariance follow
%a Wigner distribution PðsÞ ¼ ð=2Þs exp½s2=4. In

%%%%%%%%%%%%%%%%%%%%%%%%%%%%%%%%%%%%%%%%%%%%%%%%%%%%%%%%%%%%%%%%%%%%%%%%%%%%%%%%%%%%%%%%%%%%%%%%%%
\section{Conclusions}\label{conclusion}

We have characterized the ESQPT of the 3-level LMG using a wide range of indicators. Starting from the numerical calculations, the energy surface for a low number of particles $N=15$ presents clusters of levels, forecasting where the ESQPT will be found. When the dimensionality increases to $N=100$ particles, the number of level crossings increases and the ESQPT cannot be visually characterized. Hence, we make use of the mean field limit to find the ESQPT separatrices. The energy surface minimization results in a long list of stationary points with degeneration, which can be reduced to 12 different separatrices. At this point, a connection between chaos and ESQPT is made, using four of these separatrices to define the following dynamical classes: quasi-integrable, chaotic, and quasichaotic. The classification can only be made for high values of the interaction $\lambda\simeq1$, due to the lack of a wide
energy range of integrable states for the rest of $\lambda$ values. The dynamical regions are graphically represented by Poincaré sections and Peres lattices for $N\to\infty$ and $N=100$ respectively. Regular patterns are observed in the quasi-integrable regions, whereas scattered points appear in the chaotic ones.   

The numerical spectrum is also depicted with the density of states histogram, where the changes in curvature are associated with the four main ESQPT separatrices. To erase spectral fluctuations, the unfolding procedure smooths the DOS, so we can compute the nearest-neighbor spacing distribution. For high interaction $\lambda=0.98$, the spacing distributions resemble the Wigner and Poisson shapes in the chaotic and quasi-integrable regions respectively, demonstrating the utility of computing ESQPT separatrices analytically to detect chaos numerically afterwards. In addition, we provide two magnitudes that measure the chaoticity of the spacing distribution, the degree of chaos $\eta$ and the Kullback–Leibler divergence. Both present low and high values for chaotic and quasi-integrable classes respectively, with the KL divergence having the best performance and matching of the separatrices. Finally, the participation ratio and the overlap distribution are presented to detect ESQPTs and dynamical classes respectively. 

In general, the characterization of ESQPT and chaos is consistent for the limit case of the interaction parameter $\lambda\to 1$, but limited elsewhere due to two main causes: the variety of stationary points of the mean field energy surface \eqref{Separatrices} (that merge at $\lambda\simeq 1$, see Fig.\ref{Fig:Energy100} and Eq.\eqref{SeparatricesLimit}), and the impossibility to distinguish dynamical regions in a considerable energy range \cite{zenodo_Poincare_sec_videos}. For this reason, the natural extension of this work would be to compute the chaotic volume (or another quantitative chaos measure) for all $\lambda\in[0,1]$, and then check if it correlates the Morse index and hence the ESQPTs. In addition, we would like to extend the ESQPT analysis in further works to other permutation symmetry sectors of the 3-level LMG model \cite{nuestroPRE,Lieb-Mattis_PRE}, and also to the 3-level extension of other models such as the Jaynes-Cummings \cite{del_Rio-Lima_2024,ALGARNI2022106089}, Dicke \cite{PhysRevA.89.032101,PhysRevA.89.032102} or Agassi \cite{PhysRevC.97.054303,Garcia-Ramos_2019} models.

The analytical and numerical techniques developed in this work establish a systematic method applicable to other models. In fact, the study of 3-level system spectra and dynamics has a direct impact on current research topics. The applications range from novel quantum sensing devices, such as NV-centers \cite{PhysRevLett.126.220402,Lopez-Garcia2025}, to the usage of qutrits in quantum computation \cite{Cao_2024,9069177}. The traditional 2-level (qubits) framework is transforming via the extension to more general and flexible level structures, which could eventually outperform the current quantum technologies \cite{ACAR2025129404,Bottrill_2025}.

%%%%%%%%%%%%%%%%%%%%%%%%%%%%%%%%%%%%%%%%%%%%%%%%%%%%%%%%%%%%%%%%%%%%%%%%%%%%%%%%%%%%%%%%%%%%%%%%%%
\section*{Acknowledgments}
This work is part of the I+D+i projects PID2022-136228NB-C22 , and PID2023-146401NB-I00 funded by MICIU/AEI/10.13039/501100011033 and by
ERDF, EU, and of the PPIT – FEDER project SOL2024-31833 at University of Sevilla. It has also been partially supported by the Consejer\'{\i}a de Conocimiento, Investigaci\'on y Universidad, Junta de Andaluc\'{\i}a (Spain) and European Regional Development Fund (ERDF) under the grant Group FQM-160. AM thanks the support of the Spanish Ministry of Science through the project
PID2022-138144NB-I00. AS acknowledges that this work has been financially supported by the Ministry for Digital Transformation and of Civil Service of the Spanish Government through the QUANTUM ENIA project call - Quantum Spain project, and by the European Union through the Recovery, Transformation and Resilience Plan - NextGenerationEU within the framework of the Digital Spain 2026 Agenda. This work was also co-financed by the European Union, the Ministerio de Hacienda y Funci\'on P\'ublica, FEDER, and the Junta de Andaluc\'{\i}a under Project SOL2024-31833 (J.M.A., P.P-F.). %\end{acknowledgments}

\appendix

\section{$\rmu(3)$-spin coherent states}\label{statesymmatsec}

The standard $\rmu(2)$-spin coherent states are defined as a two-mode binomial expression \cite{Perelomov,Radcliffe,GilmorePhysRevA.6.2211}. $\rmu(3)$-spin coherent states (3SCSs for brevity) are defined as a three-mode multinomial generalization \eqref{cohD}. They are labeled by $2$ complex numbers $\zb=(z_1,z_2)^t\in \mathbb{C}^{2}$. In fact, we select certain patch of the complex projective manifold $\mathbb CP^{2}$, which is equivalent to choose $i=0$ as reference level \cite{QIP-2021-Entanglement}. The total number of particles $N$ is also a parameter of the 3SCS, omitted to simplify the notation. 

The coefficients $c_\nb(\zb)$ of the expansion 
of $|\psi\ra=|\zb\ra$ in the Fock basis are simply 
\be
c_\nb(\zb)=\sqrt{\frac{N!}{\prod_{i=0}^{2} n_i!}}\frac{\prod_{i=1}^{2} z_i^{n_i}}{(1+\zb^\dag\zb)^{N/2}},\label{coefCS}
\ee
where $\zb^\dag\zb=|z_1|^2+|z_{2}|^2$ denotes the standard scalar product in $\mathbb{C}^{2}$. 
%\tcb{Para ser mas consistente con la notacion usada en los articulos de Entanglement y Entropy de QIP, definiria $\zb^\dag\cdot\zb=1+|z_1|^2+\dots+|z_{D-1}|^2$, que queda mas compacto y mejor visualmente. No  me importa reescribirlo todo (Alberto)}

In general, 3SCSs are not orthogonal since the scalar product 
\be \la \zb|\zb'\ra=\frac{(1+\zb^\dag \zb')^N}{(1+\zb^\dag \zb)^{N/2}(1+\zb'^\dag \zb')^{N/2}}\label{scprod}
\ee
is not necessarily zero.  However, they are an overcomplete set of states closing a resolution of the identity \cite{nuestroPRE,QIP-2021-Entanglement, bengtsson_zyczkowski_2006}.

3SCS matrix elements of $3$-spin operators $S_{ij}$ are easily computed from \eqref{Sijmatrix} and \eqref{coefCS},
\be\label{CSEV}
\la \zb'|S_{ij}|\zb\ra=N \bar{z}'_i z_j\frac{(1+\zb'^\dag\zb)^{N-1}}{(1+\zb'^\dag\zb')^{N/2}(1+\zb^\dag\zb)^{N/2}},
\ee
where we understand $z_0=1=z'_0$. From here, 3SCS matrix elements of quadratic powers of $3$-spin operators can be concisely written as

\begin{align}\label{CSEV2}
	\langle \zb'|S_{ij}S_{kl}|\zb\rangle=&\,\delta_{jk}\la \zb'|S_{il}|\zb\ra\\
	&+\frac{N-1}{N}\frac{ \la\zb'|S_{ij}|\zb\ra\la \zb'|S_{kl}|\zb\ra}{\la \zb'|\zb\ra}\,.\nonumber
\end{align}
Note that
\begin{equation}
	\lim_{N\to\infty}\frac{ \langle \zb|S_{ij}S_{kl}|\zb\rangle}{\langle \zb|S_{ij}|\zb\rangle\langle \zb|S_{kl}|\zb\rangle}= 1,\label{nofluct}
\end{equation}
which means that quantum fluctuations are negligible in the thermodynamic limit $N\to\infty$. These ingredients are compulsory when computing the energy surface in Section \ref{sec3}.

%%%%%%%%%%%%%%%%%%%%%%%%%%%%%%%%%%%%%%%%%%%%%%%%%%%%%%%%%%%%%%%%%%%%%%%%%%%%%%%%%%%%%%%%%%%%%%%%%%%%%%

\section{Energy surface minimization: stationary points and ESQPT separatrices.}\label{App:B}

The global minimization of the energy surface \eqref{enersym},
$E^{(0)}(\lambda)=\mathrm{min}_{z_1,z_2\in \mathbb{C}}E(z_1,z_2,\lambda)$,
is achieved at the critical (real) coherent state parameters
% Stationary_points_LMG_U3.nb
\begin{eqnarray}
z_{1\pm}^{(0)}(\lambda)&=&\pm\left\{\begin{array}{lll}
	0, && 0\leq \lambda \leq \frac{1 }{3}, \\
	\sqrt{\frac{3\lambda- 1 }{ \lambda +1 }}, && \frac{1 }{3}\leq \lambda \leq \frac{3  }{5}, \\
	\sqrt{\frac{2\lambda }{-\lambda +3  }}, && \lambda \geq \frac{3  }{5},
\end{array}\right.\nonumber\\
z_{2\pm}^{(0)}(\lambda)&=&\pm\left\{\begin{array}{lll}
	0, & & 0\leq \lambda \leq  \frac{3  }{5}, \\
	\sqrt{\frac{5 \lambda -3 }{-\lambda +3  }}, & & \lambda \geq \frac{3  }{5}. \end{array}\right. \label{critalphabeta}
\end{eqnarray}
% \textcolor{blue}{In the 1st phase, $z_{10},z_{20}=0$, so it only exists the symmetric 3CAT$_{(0,0)}$. In the 2nd phase $z_{20}=0$, so there are 2 non null cats 3CAT$_{(0,0)}$, 3CAT$_{(1,0)}$. The second cat is candidate to be 1st excited state,\textcolor{red}{ future work}. In the 3rd phase all the 4 cats are present. }
Inserting \eqref{critalphabeta} into \eqref{enersym} gives the ground-state energy per particle \eqref{energysym} in the thermodynamic limit. The rest of the solutions of the equation $\nabla_{\zb} E(\zb,\lambda)=\mathbf{0}$, define the stationary points $\zb^{(j)}$ of the LMG Hamiltonian. They establish a formalism to identify and classify the ESQPTs of quantum systems 
in the classical limit \cite{CejnarReview, Stransky16}. 
We discard the solutions that have non-real coordinates in the $(q_1,q_2,p_1,p_2)$ space (see Eq.(16)) for all
$\lambda$, or violate the norm constraint $q_1^2+q_2^2+p_1^2+p_2^2\leq 2$.
The potential ESQPT separatrices are defined as energy surface evaluated in the stationary points $f_j(\lambda)=E(\zb^{(j)},\lambda)$. Note that this definition is patch-dependent, as it is linked to the coherent states definition \eqref{cohD} in the patch $i=0$ of the complex projective manifold $\mathbb{C}P^2$. This produces a limitation in the $\lambda$-domain of definition of the separatrices, which is studied in depth in App.\ref{App:C}. For the sake of simplicity, we enlist below the patch-independent version of the separatrices with the extended $\lambda$-domain,
%Stationary_points_LMG_U3.nb/Domain of the other separatrices

\begin{align}\label{Separatrices}
	\begin{array}{llllr}
		f_{0}(\lambda)=1-\frac{1}{2\lambda}-\frac{7\lambda}{6},  && \forall \lambda \geq  \frac{3}{5}, && (E_0)\\
        f_{1}(\lambda)=-\frac{(1+ \lambda )^2}{8 \lambda }, && \forall \lambda \geq \frac{1 }{3}, &&  \mathrm{(ESQPT 1)}\\
        f_2(\lambda)=-1+\lambda,  && \forall \lambda \geq 0, && \mathrm{(ESQPT 2)}\\
        f_3(\lambda)=\frac{(1-3\lambda)^2}{10\lambda},  && \forall \lambda \geq  \frac{1}{3}, && \mathrm{(ESQPT 3)}\\
        f_{4}(\lambda)=\frac{(1+\lambda )^2}{8 \lambda },  && \forall \lambda \geq  \frac{1}{3}, && (E_{\text{max}})\\
        f_{5}(\lambda)=-\frac{(1- 3\lambda )^2}{8 \lambda },  && \forall \lambda \geq  \frac{1}{3}, && \\
        f_6(\lambda)=1-\frac{1}{2\lambda}-\lambda,  && \forall \lambda \geq  \frac{1}{2}, && \\
        f_{7}(\lambda)=0,  && \forall \lambda \geq  0, && \\
        f_8(\lambda)=1-\lambda,  && \forall \lambda \geq  0, && \mathrm{(ESQPT 8)}\\
        f_{9}(\lambda)=1-\frac{3+\lambda^2}{6 \lambda },  && \forall \lambda \geq  \frac{3}{4}, && \mathrm{(ESQPT 9)}\\ 
        f_{10}(\lambda)=\frac{(1+\lambda)^2}{10\lambda},  && \forall \lambda \geq  \frac{2}{3}, && \\
        f_{11}(\lambda)=1-\frac{5+\lambda^2}{10\lambda},  && \forall \lambda \geq  \frac{5}{7}, && \\
        f_{12}(\lambda)=\frac{(1-3\lambda)^2}{8\lambda},  && \forall \lambda \geq  \frac{1}{3}, && \\
        f_{13}(\lambda)=-1+\frac{1}{2\lambda}+\lambda,  && \forall \lambda \geq  \frac{1}{2}. && \\  
	\end{array}
\end{align}

\begin{figure}[h]
%Retrieved from Stationary_points_LMG_U3_v2.nb
	\centering
	\includegraphics[width=0.45\textwidth]{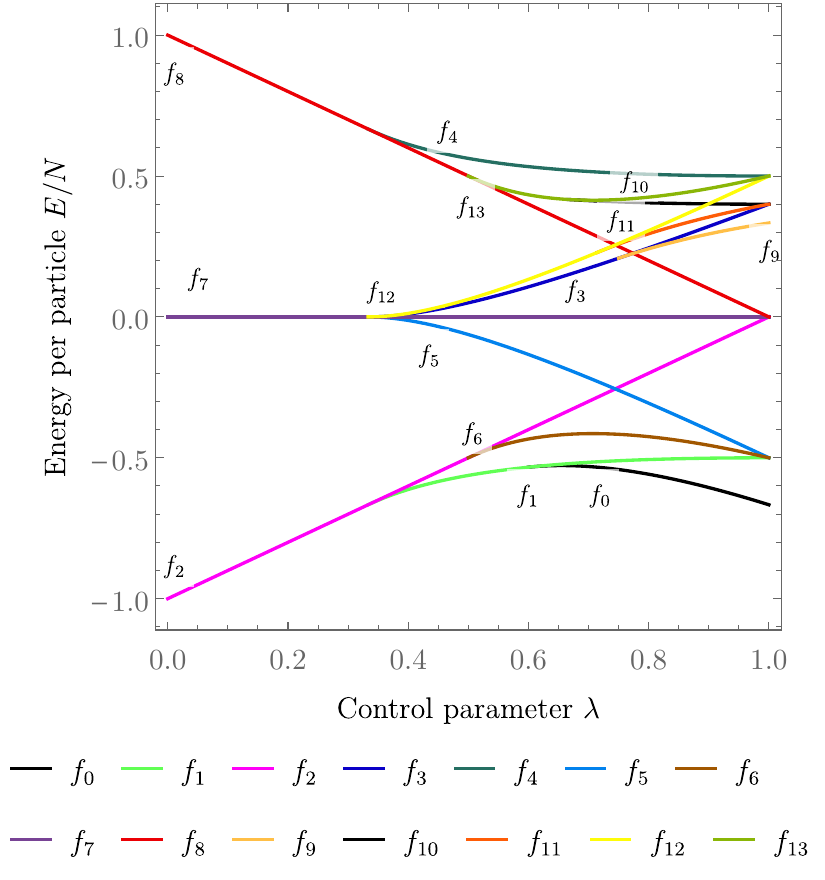}
	\caption{Separatrices defined as energy surfaces evaluated in the stationary points $f_j(\lambda)=E(\zb^{(j)},\lambda)$ in eq.\eqref{Separatrices}.}
	\label{Fig:SeparatricesOnly}
\end{figure}

\begin{table}[H]
%Retrieved from Stationary_points_LMG_U3_v2.nb
$$
\begin{array}{|c|c|c|c|c|c|c|c|}
\hline
&
\left[0,\frac{1}{3}\right] &
\left[\frac{1}{3},\frac{1}{2}\right] &
\left[\frac{1}{2},\frac{3}{5}\right] &
\left[\frac{3}{5},\frac{2}{3}\right] &
\left[\frac{2}{3},\frac{5}{7}\right] &
\left[\frac{5}{7},\frac{3}{4}\right] &
\left[\frac{3}{4},1\right]
\\
\hline

r_0
& & & &
0 & 0 & 0 & 0
\\
\hline

r_1
& & 0 & 0 &
{\color{green}1} & {\color{green}1} & {\color{green}1} & {\color{green}1}
\\
\hline

r_2
& 0 & {\color{green}1} &
{\color{magenta}2} &
{\color{magenta}2} &
{\color{magenta}2} &
{\color{magenta}2} &
{\color{magenta}2}
\\
\hline

r_3
& & {\color{magenta}2} &
{\color{magenta}2} &
{\color{magenta}2} &
{\color{magenta}2} &
{\color{magenta}2} &
{\color{blue}3}
\\
\hline

r_4
& & {\color{green}1} &
{\color{magenta}2} &
{\color{blue}3} &
{\color{blue}3} &
\textcolor{darkbrown}{4}  &
\textcolor{darkbrown}{4} 
\\
\hline

r_5
& & {\color{green}1} &
{\color{green}1} &
{\color{green}1} &
{\color{green}1} &
{\color{green}1} &
{\color{green}1}
\\
\hline

r_6
& & &
{\color{green}1} &
{\color{green}1} &
{\color{green}1} &
{\color{green}1} &
{\color{green}1}
\\
\hline

r_7
& {\color{magenta}2} &
{\color{magenta}2} &
{\color{magenta}2} &
{\color{magenta}2} &
{\color{magenta}2} &
{\color{magenta}2} &
{\color{magenta}2}
\\
\hline

r_8
& \textcolor{darkbrown}{4}  &
{\color{blue}3} &
{\color{magenta}2} &
{\color{magenta}2} &
{\color{magenta}2} &
{\color{magenta}2} &
{\color{magenta}2}
\\
\hline

r_9
& & & & & & &
{\color{magenta}2}
\\
\hline

r_{10}
& & & & &
{\color{blue}3} &
{\color{blue}3} &
{\color{blue}3}
\\
\hline

r_{11}
& & & & & &
{\color{blue}3} &
{\color{blue}3}
\\
\hline

r_{12}
& & {\color{blue}3} &
{\color{blue}3} &
{\color{blue}3} &
{\color{blue}3} &
\textcolor{darkbrown}{4}  &
\textcolor{darkbrown}{4} 
\\
\hline

r_{13}
& & &
{\color{blue}3} &
{\color{blue}3} &
\textcolor{darkbrown}{4}  &
\textcolor{darkbrown}{4}  &
\textcolor{darkbrown}{4} 
\\
\hline

\end{array}
$$
    \caption{Indexes $r_j$ of the stationary points in eq.\eqref{StatPoints} at different regions of the control parameter $\lambda$. Each color represents a different value of the indixes.}
    \label{tab:r}
\end{table}

The separatrices are represented isolated, without the  energy spectrum for a finite number of particles, in Figure \ref{Fig:SeparatricesOnly}. We can observe how $f_2$, $f_1$, $f_0$ define the ground state of the semiclassical model in the regions $\lambda\in[0,1/3]$, $\lambda\in[1/3,3/5]$, and $\lambda\in[3/5,1]$ respectively. Interestingly, the separatrices form clusters in the strong interaction limit,
\begin{align}\label{SeparatricesLimit}
    \begin{array}{llr}
    \lim_{\lambda\to 1} f_{0}=-\frac{2}{3}, &&  \mathrm{(lowest}\:\mathrm{energy)} \\
    \lim_{\lambda\to 1} f_1,f_5,f_{6}=-\frac{1}{2}, &&  \mathrm{(ESQPT 1)} \\
    \lim_{\lambda\to 1} f_2,f_7,f_{8}=0, &&  \mathrm{(ESQPT 2,8)}\\
    \lim_{\lambda\to 1} f_9=\frac{1}{3}, && \mathrm{(ESQPT 9)} \\
    \lim_{\lambda\to 1} f_3,f_{10},f_{11}=\frac{4}{10}, &&  \mathrm{(ESQPT 3)}\\
    \lim_{\lambda\to 1} f_4,f_{12}, f_{13}=\frac{1}{2}. && \mathrm{(highest}\:\mathrm{energy)} \\
    \end{array}
\end{align}

These limits inspire us to define the dynamical regions in \eqref{DynamicalRegions}. We must remark that this link between the semiclassical model (ESQPT) and the chaotic regions is only valid in the $\lambda\to 1$ as previously stated in \cite{Meredith}, considering that a general connection between ESQPT and chaos does not exist \cite{STRANSKY201557,Cejnar_2015}.

The stationary points $\bm{z}^{(j)}$ are expressed in the $(q_1^{(j)},q_2^{(j)},p_1^{(j)},p_2^{(j)})$ variables (see eqs.\eqref{changeCoor1},\eqref{changeCoor2}), where the minimization of the energy surface is easier. Note that these expressions are $i=0$ patch dependent. That is, to obtain the extended $\lambda$-domain of eq.\eqref{Separatrices}, another patches must be considered. See App.\ref{App:C} for more details. In addition, the stationary point of the solution $f_{4}$ ($E_{\text{max}}$) is not achievable in the patch $i=0$, but in $i=2$, hence this stationary point shown in the following equation corresponds to the $i=2$ patch coordinates $(q_0^{(j)},q_1^{(j)},p_0^{(j)},p_1^{(j)})$ as an exception.
\begin{align}\label{StatPoints}
	\begin{array}{llllr}
        (\pm\sqrt{\frac{2}{3}},\pm\sqrt{\frac{-3+5\lambda}{3\lambda}},0,0) && \forall \lambda \geq \frac{3}{5}, && (f_{0})\\
        (\pm\sqrt{\frac{-1+3\lambda}{2\lambda}},0,0,0) && \forall \lambda \geq \frac{1}{3}, && (f_1)\\
        (0,0,0,0) && \forall \lambda \geq 0, && (f_2)\\
        (0,0,\pm\sqrt{\frac{4-2\lambda}{5\lambda}},\pm\sqrt{\frac{-1+3\lambda}{5\lambda}}) && \forall \lambda \geq \frac{1}{3}, && (f_3)\\
        (0,0,0,\pm\sqrt{\frac{-1+3\lambda}{2\lambda}}) && \forall \lambda \geq \frac{1}{3}, && (f_{4})\\
        (\pm\frac{\sqrt{3(-1+3\lambda)(1+\lambda)}}{4\lambda},
        \pm\frac{\sqrt{3(-1+3\lambda)^2}}{4\lambda}, && && \\
        \:\:\:\pm\frac{\sqrt{(3-\lambda)(1+\lambda)}}{4\lambda},
        \pm\frac{\sqrt{(-1+3\lambda)(3-\lambda)}}{4\lambda} )
        && \forall \lambda \geq \frac{1}{3}, && (f_{5})\\
        (0,\pm\sqrt{\frac{-1+2\lambda}{\lambda}},0,0) && \forall \lambda \geq \frac{1}{2}, && (f_6)\\
        (\pm\sqrt{\frac{-1+3\lambda}{2\lambda}},0,\pm\sqrt{\frac{1+\lambda}{2\lambda}},0) && \forall \lambda \geq \frac{1}{3}, && (f_{7})\\
        (0,\pm\sqrt{\frac{-1+2\lambda}{\lambda}},0,\pm\frac{1}{\sqrt{\lambda}}) && \forall \lambda \geq \frac{1}{2}, && (f_8)\\
        (\pm \sqrt{\frac{(-3+4\lambda)}{3(3-\lambda)}},
        \pm\sqrt{\frac{(3-2\lambda)(-3+4\lambda)}{3\lambda(3-\lambda)}}, && && \\
        \:\:\:\sqrt{\frac{(3-2\lambda)}{(3-\lambda)}},
        \pm\sqrt{\frac{\lambda(3-4\lambda)(-3+2\lambda)}{\lambda(-3+\lambda)(9+2\lambda(-9+4\lambda))}} )
        && \forall \lambda \geq \frac{3}{4}, && (f_{9})\\
        (\pm\sqrt{\frac{2}{5}}\sqrt{\frac{-2+3\lambda}{\lambda}},0,0,\pm\sqrt{\frac{3}{5}}\sqrt{\frac{1+\lambda}{\lambda}}) && \forall \lambda \geq \frac{2}{3}, && (f_{10})\\
        (0,\pm\sqrt{\frac{-5+7\lambda}{5\lambda}},\pm\sqrt{\frac{6}{5}},0) && \forall \lambda \geq \frac{5}{7}, && (f_{11})\\
        (0,0,\pm\sqrt{\frac{1+\lambda}{2\lambda}},0) && \forall \lambda \geq \frac{1}{3}, && (f_{12})\\
        (0,0,0,\pm\frac{1}{\sqrt{\lambda}}) && \forall \lambda \geq \frac{1}{2}. && (f_{13})\\
        % \frac{\sqrt{(-1+3\lambda)}}{4\lambda}
        % (\scriptstyle{\pm\sqrt{3(1+\lambda)},
        % \pm\sqrt{3(-1+3\lambda)},
        % \pm\frac{\sqrt{(3-\lambda)(1+\lambda)}}{(-1+3\lambda)},
        % \pm\sqrt{(3-\lambda)})
        % && \forall \lambda \geq \frac{1}{3}, && (f_{13})\\
	\end{array}
\end{align}

The plus-minus signs indicate the $ 1, 2, 4$, or $8$-degeneracy of the points. That is, the signs are independent with all combinations allowed, with the exception of $f_5$ where only an even number of $+$ signs are allowed, and $f_9$ where only an odd number of $+$ are allowed. We can classify the stationary point by computing the eigenvalues of the Hessian matrix $\partial_{z_m}\partial_{z_n}E(\bm{z},\lambda)\Bigr|_{\substack{\bm{z}=\bm{z}^{(j)}}}$ \cite{CejnarReview}. For each stationary point, we obtain 4 eigenvalues, whose sign depend on the control parameter $\lambda$. When the point $\bm{z}^{(j)}(\lambda)$ has not any null eigenvalue (for a fixed $\lambda$), we call it non-degenerate, and we define the index $r_j$ as the number of negative eigenvalues that this stationary point has. 
This behavior is shown in Table \ref{tab:r}, taking into account the $\lambda$-domains in eqs.\eqref{Separatrices},\eqref{StatPoints}. The separatrices with $r=0$ define the local minimum or ground state, a combination of $f_2$, $f_1$, and $f_0$; the separatrices with $r=4$ are local maxima ($f_8$ and $f_4$), and the intermediate values of $r$ define multidimensional saddle points. We can also observe how the indexes change their values with the different colors of the numbers, identifying the values of $\lambda$ where some stationary points are degenerate. Some of these control parameter values coincide with the QPT at $\lambda=\frac{1}{3},\frac{3}{5}$.

\begin{figure}[h]
%Retrieved from Stationary_points_LMG_U3_v2.nb
	\centering
    \includegraphics[width=0.45\textwidth]{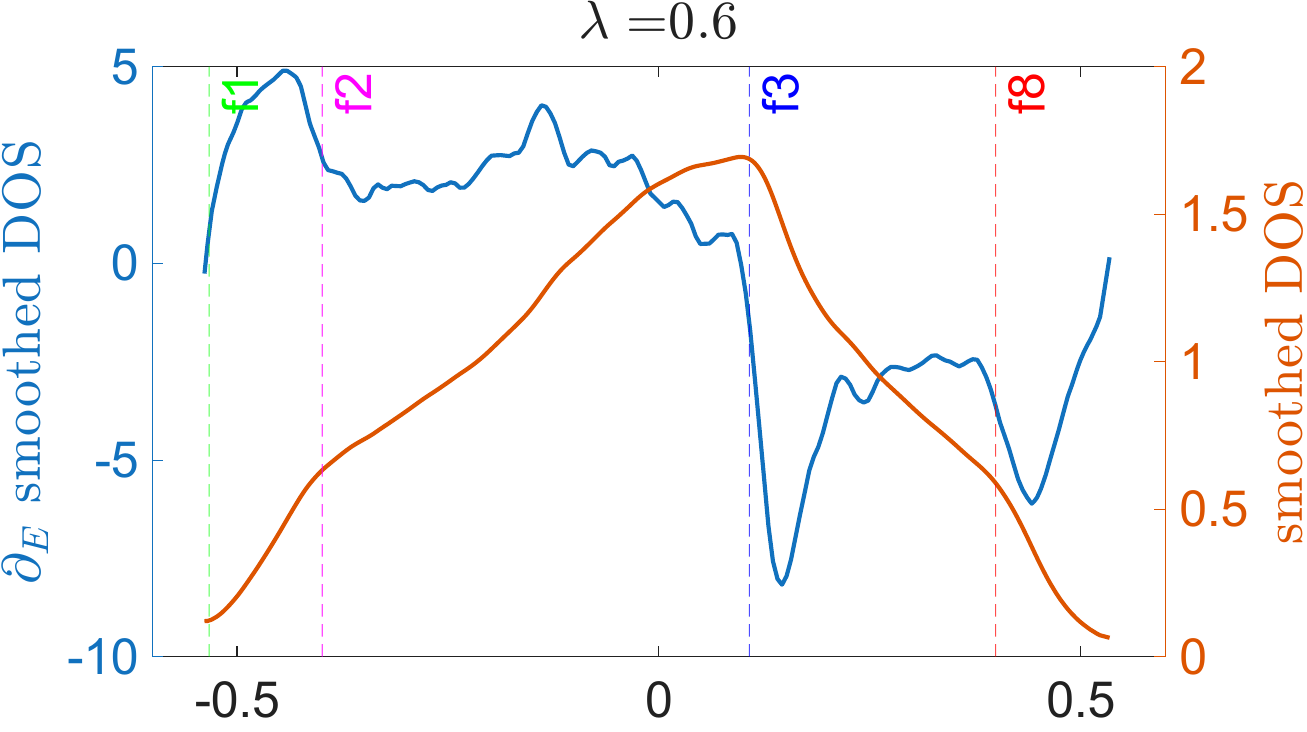}
	\includegraphics[width=0.45\textwidth]{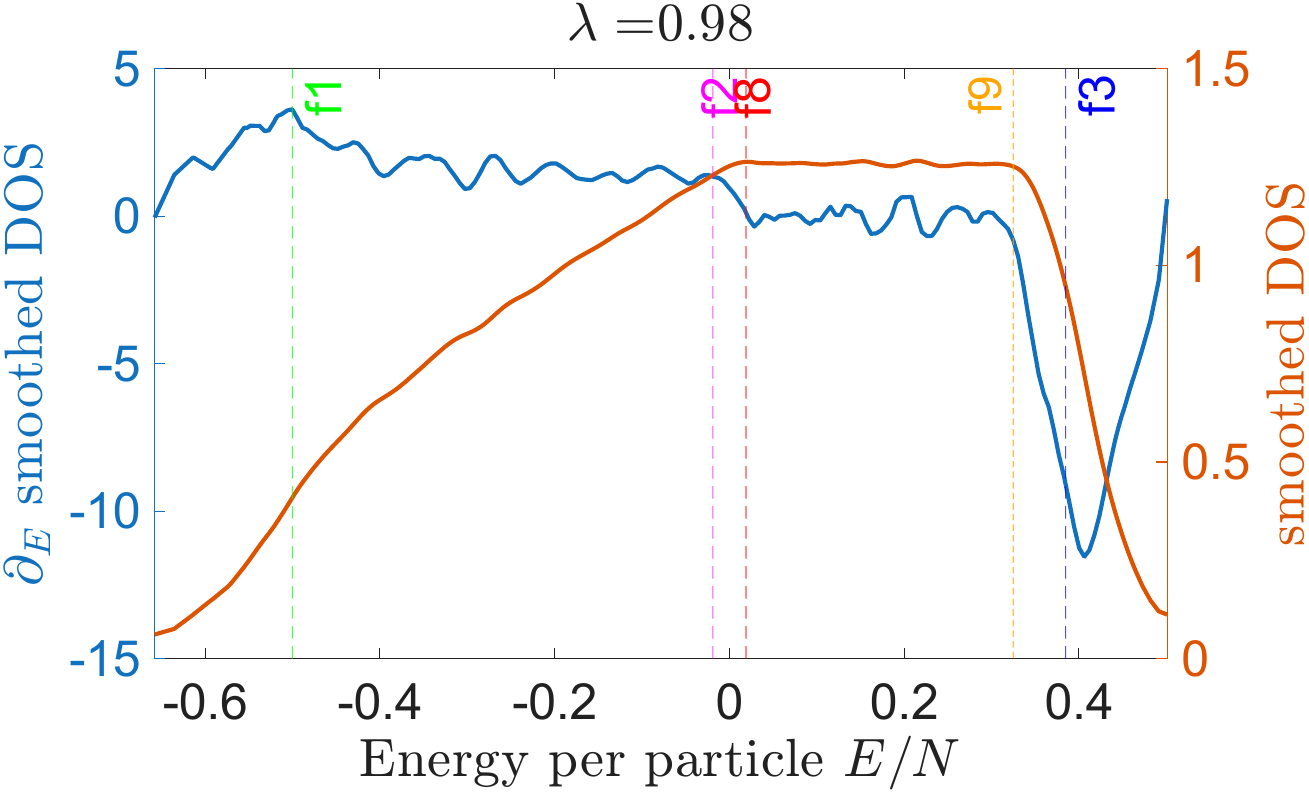}
	\caption{Smoothed density of states (right) and its derivative (left) as functions of the energy per particle of the 3-level LMG model, for $N=100$ particles and two values of the control parameter $\lambda=0.6,0.98$. The spectrum was unfolded over a DOS histogram of 200 bins, smoothed by a Gaussian filter with a 15-elements window \cite{Haake}. The dashed vertical lines represent the ESQPT separatrices \eqref{Separatrix1}, and \eqref{Separatrix2}.   }
	\label{Fig:UnfoldedDOSderivative}
\end{figure}

A non-degenerate stationary point of index $r$ at a specific energy alters the smoothed density of states according to the Morse lemma \cite{Stransky16}. More precisely, in a system with 2 degrees of freedom, as the 3-level LMG, the stationary points induce discontinuities in the first derivative of the density of states, which can be a jump ($r$ even) or a logarithmic ($r$ odd) divergence. We have computed the first and the second derivatives of the integral of the smoothed DOS $\sigma_{\Delta}(E)$ (previously computed in Sec. \ref{Sec:NNSD}), which are equivalent to the smoothed version of the DOS and its derivative, respectively. In particular, we only plot them in Figure \ref{Fig:UnfoldedDOSderivative} for two representative values of the control parameter $\lambda=0.6$, and $0.98$ where all the selected separatrices appear. In the end, the unfolded DOS derivative shows jump divergences for the separatrices in both panels of Figure \ref{Fig:UnfoldedDOSderivative}, according to the classification of Table \ref{tab:r}. For $\lambda=0.6$ (top panel), the discontinuity in the derivative of $f_2,f_3,f_8$ seems to be a jump, according to the indexes equal to 2 given in Table \ref{tab:r}, whereas $f_1$ is transitioning from a local minimum $r_1=0$ to a logarithmic divergence $r_1=1$. In contrast, for $\lambda=0.98$ (bottom panel), the divergence of $f_3$ seems to tend to a logarithmic shape, as $r_3=3$ in this region. The rest of the separatrices indexes remain the same at $\lambda=0.98$, in addition to the separatrix $f_9$ that appears at this $\lambda$ value with a $r_9=2$ jump divergence.

Despite the given results, Figure \ref{Fig:UnfoldedDOSderivative} does not provide an exact criterion for classifying ESQPT in contrast to other models in the literature \cite{CejnarReview}. This is mainly due to the noise that presents the DOS histogram (see Figure \ref{Fig:DOSmultiple100}) for a chaotic system. Nevertheless, from \ref{tab:r}, we obtain the positive result that the Morse index correlates with relevance to chaos at $\lambda\simeq 1$. 

%%%%%%%%%%%%%%%%%%%%%%%%%%%%%%%%%%%%%%%%%%%%%%%%%%%%%%%%%%%%%%%%%%%%%%%%%%%%%%%%%%%%%%%%%%%%%%%%%%%%%%

\section{$\lambda$-domain of the separatrices and patch selection of the complex projective $\mathbb{C}P^2$}\label{App:C}

In Appendix \ref{statesymmatsec} we fixed the level $i=0$ as a reference to define the 3SCSs, which is equivalent to selecting a certain patch of the complex projective manifold $\mathbb CP^{2}$. The problem comes when we try to reach energy levels accessible only in the limits $|z_1|,|z_2|\to\infty$, which are out of the domain of the patch coordinates. To be mathematically precise, the $\|\bm{z}\|\to\infty$ limit does not exist since its value depends on the direction, see \cite{Guerrero_2023} for an example using hyper-spherical coordinates in a 3-level LMG model. For instance, if we try to obtain the highest energy level, i.e. the separatrix $f_8=(1-\lambda)\:\:\forall\lambda\geq 0$ \eqref{Separatrices}, we have to take the limits $|z_1|\to0,|z_2|\to\infty$ in the energy surface expression \eqref{enersym}. To prevent the limit, another patch should be chosen. That is, we should take the level $i=2$ as reference, redefine the 3SCSs, recalculate the matrix elements \eqref{CSEV} and the energy surface, and compute the separatrices in the new patch.\\

All these calculations are included in the supplemental material \cite{zenodo_statpoints}, where we compute the separatrices in two different new patches, with $i=1$ and $i=2$ as reference levels. The rest of separatrices remains the same, but for the domain. 
The separatrices obtained are the same as in the $i=0$ patch, with the difference that the $\lambda$-domains of $f_{8}$ and $f_{10}$, originally $\lambda\geq 1/3$ and $\lambda\geq 1/2$, are extended to $\lambda\geq0$ in the $i=1$ and $i=2$ patches, respectively. The price to pay is that other separatrices, as $f_2$, have a reduced domain in these patches compared to the $i=0$ case. It is also worth mentioning that the separatrix $f_{7}$ has an extended domain $\lambda\geq 0$ when computed in the $i=2$, however, their respective stationary point $(\pm\sqrt{(1+\lambda)/2\lambda},0,0,0)$ does not fulfill the conservation criterion $q_1^2+q_2^2+p_1^2+p_2^2\leq 2$ for $\lambda<1/3$, hence its $\lambda$-domain remains the same.\\

In short, the selection of a different patch makes a more complete visualization of the separatrices limit in Figure \ref{Fig:Energy100}. That is, the highest energy state converges with $f_8$ even for low values of the control parameter, and the energy 0 state is equivalent to the $f_{7}$ separatrix.

%%%%%%%%%%%%%%%%%%%%%%%%%%%%%%%%%%%%%%%%%%%%%%%%%%%%%%%%%%%%%%%%%%%%%%%%%%%%%%%%%%%%%%%%%%%%%%%%%%%%%%

\appendix

%%%%%%%%%%%%%%%%%%%%%%%%%%%%%%%%%%%%%%%%%%%%%%%%%%%%%%%%%%%%%%%%%%%%%%%%%%%%%%%%%%%%%%%%%%%%%%%%%%%%%%

%\bibliographystyle{apsrev4-1}
%\bibliography{/home2/comun/MEGA/Genfimat/Bibliografia/bibliografia.bib}
%\bibliography{/home/usuario/MEGA_Genfimat/Bibliografia/bibliografia.bib}
%\bibliography{/media/DATOS/calixto/MEGA_Genfimat/Bibliografia/bibliografia.bib}
%\bibliography{/home2/comun/MEGA/Genfimat/Bibliografia/bibliografia.bib}
\bibliography{bibliografia.bib}

\end{document}